\RequirePackage{fix-cm}
\documentclass[smallextended]{svjour3}       
\smartqed  
\usepackage{graphicx,natbib,amssymb}
\journalname{Celestial Mechanics \& Dynamical Astronomy}

\def\OmegaG{{\Omega_{\rm G}}}
\def\rhoG{{\rho_{\rm G}}}

\begin{document}

\title{Planetary Orbital Equations in Externally-Perturbed Systems:
Position and Velocity-Dependent Forces
}

\titlerunning{Perturbed Two-Body Problem}        

\author{Dimitri Veras         \and
        N. Wyn Evans 
}


\institute{Dimitri Veras \at
              Institute of Astronomy, University of Cambridge, Madingley Road, Cambridge CB3 0HA \\
              Tel.: +44 (0)1223 337548\\
              Fax: +44 (0)1223 337523\\
              \email{veras@ast.cam.ac.uk}           
           \and
           N. Wyn Evans \at
              Institute of Astronomy, University of Cambridge, Madingley Road, Cambridge CB3 0HA \\
              \email{nwe@ast.cam.ac.uk}           
}

\date{Received: 02 July 2012 / Revised: 27 September 2012 / Accepted: 24 October 2012}

\maketitle

\begin{abstract}
The increasing number and variety of extrasolar planets illustrates
the importance of characterizing planetary perturbations.  Planetary
orbits are typically described by physically intuitive orbital
elements.  Here, we explicitly express the equations of motion of the
{\bf \it unaveraged} perturbed two-body problem in terms of planetary orbital
elements by using a generalized form of Gauss' equations.  We consider
a varied set of position and velocity-dependent perturbations, and
also derive relevant specific cases of the equations: when they
are averaged over fast variables (the ``adiabatic'' approximation), and in
the prograde and retrograde planar cases.  In each instance, we delineate
the properties of the equations. As brief demonstrations of
potential applications, we consider the effect of Galactic tides.
We measure the effect on the widest-known exoplanet orbit, 
Sedna-like objects, and distant scattered disk objects, particularly 
with regard to where the adiabatic approximation breaks down. 
The {\it Mathematica} code which can help derive the equations of 
motion for a user-defined perturbation is freely available upon request.
\end{abstract}

\section{Introduction}

The discovery of extrasolar planets challenged our understanding of
the formation and subsequent evolution of planetary systems.  The
variety of dynamical architectures demonstrated by pulsar planets
\citep[e.g.][]{wolfra1992}, Hot Jupiters \citep[e.g.][]{mayque1995},
wide-orbit planets
\citep[e.g.][]{kuzetal2011,luhetal2011}, free-floating
planets \citep[e.g.][]{lucroc2000,zapetal2000,sumetal2011}, highly
inclined and retrograde planets
\citep[e.g.][]{andetal2010,trietal2010,winetal2010,kaietal2011},
closely-packed planets \citep[e.g.][]{liseatl2011}, circumbinary
planets \citep[e.g.][]{sigetal2003,doyetal2011} and the Solar System defy a
single, simple explanation for their formation.  

After formation, these systems are subjected to internal forces, from
planets and smaller bodies, and external forces, such as the singular
or repeated local close encounters with individual passing stars, the
Galactic tide, tidal tails, molecular clouds, globular clusters or
even dark matter substructures.  Further, a star's passage into and
out of spiral arms, and the deformation of the Galactic tide due to
close encounters or collisions with other galaxies can affect orbiting
planets\footnote{The Milky Way and Andromeda will collide in $\sim$ 2-5
  Gyr, before many Galactic stars, including the Sun, turn off of the
  main-sequence \citep{coxloe2008,vdm2012}.}.  One known exoplanet is
thought to be of extragalactic origin, perhaps originating from a
disrupted satellite galaxy \citep{setetal2010}.  The vast population
of free-floating planets -- nearly two per main sequence star
\citep{sumetal2011} -- which cannot be explained by planet-planet
scattering alone \citep{verray2012}, perhaps demonstrates that
external disruption of planetary systems is an endemic feature of the
Galactic disk.

Some of these external forces may be modeled as perturbations on a
two-body star-planet system.  Although the two-body problem is
solvable analytically, the perturbed two-body problem is generally not
so.  Nevertheless, expressing these equations of motion entirely in
terms of planetary orbital elements garners intuition for how
planets are affected by external perturbations and helps one obtain
analytical solutions in limiting cases.  Here, we present a method for
performing this conversion for arbitrary perturbations.

Suppose the equations of motion for a single planet orbiting a single
star are subject to extra accelerations $\Upsilon$.  Then the
equations of motion can be expressed as:

\begin{eqnarray}
\frac{d^2x}{dt^2} &=& - \frac{G \left(m_{\star} + m_p \right) x}{\left(x^2 + y^2 + z^2\right)^{3/2}} 
+ \Upsilon_{xx} x + \Upsilon_{xy} y + \Upsilon_{uu} \frac{dx}{dt} + \Upsilon_{uv} \frac{dy}{dt}
\label{xeq}
\\
\frac{d^2y}{dt^2} &=& - \frac{G \left(m_{\star} + m_p \right) y}{\left(x^2 + y^2 + z^2\right)^{3/2}} 
+ \Upsilon_{yx} x + \Upsilon_{yy} y + \Upsilon_{vu} \frac{dx}{dt} + \Upsilon_{vv} \frac{dy}{dt}
\label{yeq}
\\
\frac{d^2z}{dt^2} &=& - \frac{G \left(m_{\star} + m_p \right) z}{\left(x^2 + y^2 + z^2\right)^{3/2}} 
+ \Upsilon_{zz} z + \Upsilon_{ww} \frac{dz}{dt}
\label{zeq}
\end{eqnarray}
where $m_{\star}$ and $m_p$ represent the masses of the star and
planet, respectively.  In the context of a single planet orbiting the
Galactic center, we may take ($x,y,z$) to be a non-rotating rectangular
coordinate system centered on the star, as in \citet{heitre1986}.  The
primary goal of this paper is to express Eqs. (\ref{xeq})-(\ref{zeq})
analytically in terms of the time evolution of planetary
orbital elements.  The secondary goal is to present the method by
which one can repeat the procedure for other perturbational forms not
included in Eqs. (\ref{xeq})-(\ref{zeq}).

In Section 2, we provide background for the perturbed two-body problem
and present the algorithm used to derive the equations.  We then
present the general equations in Section 3.  Sections 4 and 5
specialise to the adiabatic and planar adiabatic cases, for which
substantial simplifications are possible. We provide applications to
perturbations derived from the Galactic tides in Section 6 and
summarize our results in Section 7.

\section{Background and Derivation Algorithm}

In order to describe fully the position and velocity of a planet, 6
orbital elements must be employed.  Four often-used elements are the
semimajor axis, $a$, eccentricity, $e$, inclination, $i$, and the
longitude of ascending node, $\Omega$.  A fifth orbital element
usually includes the longitude of pericenter, $\varpi$, or the
argument of pericenter, $\omega$.  The difference is important here,
and they are related through $\omega + \Omega = \varpi$.  All
five of these elements fully describe the shape of a planet moving on
a Keplerian ellipse.  In the absence of any other planets or forces,
these elements remain fixed, as in the classic two-body problem.  The
sixth element, which provides the location of the planet along its
orbit, can be described by one of several different measures,
including the mean anomaly $M$, true anomaly $f$, eccentric anomaly
$E$, true longitude $\theta$, mean longitude $\lambda$, or argument of
latitude $\upsilon$.  We adopt $f$ as our sixth element due to its
intuitive geometric interpretation and with an eye towards future
studies that might wish to focus on planets residing at locations
close to either pericenter or apocenter.

The perturbed two-body problem has been a subject of extensive study
and is applicable to several fields of astrophysics.  The work of
\cite{burns1976}, subsequently popularised by
\citet[][pp. 54-57]{murder1999}, provides a mechanism to obtain
analytic equations for $da/dt$, $de/dt$, $di/dt$, and $d\Omega/dt$
arising from a small perturbative force with a given prescription for
radial, tangential and normal components. This line of attack can be
traced back ultimately to Gauss (see e.g., section 9.13 of Brouwer \&
Clemence 1961).

However, there is an alternative.  Recently, a re-analysis of the
derivation of Lagrange's Planetary Equations, which describe the
perturbation from a third body, revealed that a generalized
gauge may be adopted in the derivation.  The consequences and
extensive description of this ``generalized gauge theory'' is provided
in
\cite{efrgol2003,efrgol2004}, \cite{gurfil2004}, \cite{efroimsky2005a,efroimsky2005b,efroimsky2006},
\cite{gurfil2007}
and \cite{gurbel2008}.  This theory also yields analytic equations for
$da/dt$, $de/dt$, $di/dt$, and $d\Omega/dt$ but also directly for
$dM/dt$, arising from a perturbative force with a given prescription
for each Cartesian component.  In principle, either approach can be
used for the purposes of this paper; we use the latter to derive the
equations of motion due to its generality and compactness.

To derive the equations of motion in orbital elements for {\it any}
perturbing acceleration, we choose the following form, similar to
Eq. (22) of \cite{efroimsky2005a} and Eq. (16) of \cite{gurfil2007}:

\begin{equation}
\frac{d\vec{\beta}}{dt} = \mathcal{M}_1
\left[
\mathcal{M}_2 \left( \Delta \vec{A} - \frac{d\vec{\Phi}}{dt}  \right)
-
\mathcal{M}_3 \vec{\Phi}
\right]
\label{geneq}
\end{equation}
where $\vec{\beta} =
\left\lbrace{a,e,i,\Omega,\omega,M_0}\right\rbrace$, the subscript
``0'' refers to the initial value, and $\vec{\Phi}$ is the gauge.  

The $\mathcal{M}$ represent matrices that are expressed entirely in terms of
orbital elements, and $\Delta \vec{A}$ is the acceleration caused by a
perturbative force on the system.  Although $\Delta \vec{A}$ is
typically written and referred to as a force, the quantity actually
represents an acceleration.  Equation (\ref{geneq}) is useful because
all three matrices are system-independent, and can be precomputed
before any perturbative force or gauge is applied.

Assume $\vec{A}$ is expressed as a column vector.  The matrix
$\mathcal{M}_1$ is then the transpose of the ``Poisson Matrix'', which
is composed of Poisson Brackets, and is the negative inverse of the
``Lagrange Matrix'', which is composed of the Lagrange Brackets.  This
matrix reads:

\[ \mathcal{M}_1=  \left( 
\begin{array}{cccccc}
0 & 0 & 0 & 0 & 0 & \frac{2}{na}\\
0 & 0 & 0 & 0 & -\frac{\sqrt{1-e^2}}{nea^2} & \frac{1-e^2}{nea^2}\\
0 & 0 & 0 & -\frac{1}{na^2\sin{i} \sqrt{1-e^2}} & \frac{1}{na^2\tan{i} \sqrt{1-e^2}} & 0 \\
0 & 0 & \frac{1}{na^2\sin{i} \sqrt{1-e^2}} & 0 & 0 & 0\\
0 & \frac{\sqrt{1-e^2}}{nea^2} & -\frac{1}{na^2\tan{i}\sqrt{1-e^2}} & 0 & 0 & 0\\
-\frac{2}{na} & -\frac{1-e^2}{nea^2} & 0 & 0 & 0 & 0
\end{array} 
\right)\] 

\noindent{where} the orbital elements associated with each row
starting at the top are $a$, $e$, $i$, $\Omega$, $\omega$ and $M_0$,
respectively. Note that these matrix entries represent the
coefficients in the appropriate form of Lagrange's Planetary
Equations.  The matrices $\mathcal{M}_2$ and $\mathcal{M}_3$ are
composed of partial derivatives of $\vec{r}$ and $\vec{v}$, namely:
\[ \mathcal{M}_2=  \left( 
\begin{array}{cccccc}
\frac{\partial r_x}{\partial a} & \frac{\partial r_x}{\partial e}  & \frac{\partial r_x}{\partial i} & \frac{\partial r_x}{\partial \Omega} & \frac{\partial r_x}{\partial \omega} & \frac{\partial r_x}{\partial M_0}\\
\frac{\partial r_y}{\partial a} & \frac{\partial r_y}{\partial e} & \frac{\partial r_y}{\partial i} & \frac{\partial r_y}{\partial \Omega} & \frac{\partial r_y}{\partial \omega} & \frac{\partial r_y}{\partial M_0}\\
\frac{\partial r_z}{\partial a} & \frac{\partial r_z}{\partial e} & \frac{\partial r_z}{\partial i} & \frac{\partial r_z}{\partial \Omega} & \frac{\partial r_z}{\partial \omega} & \frac{\partial r_z}{\partial M_0} 
\end{array} 
\right)\] 
\[ \mathcal{M}_3=  \left( 
\begin{array}{cccccc}
\frac{\partial v_x}{\partial a} & \frac{\partial v_x}{\partial e}  & \frac{\partial v_x}{\partial i} & \frac{\partial v_x}{\partial \Omega} & \frac{\partial v_x}{\partial \omega} & \frac{\partial v_x}{\partial M_0}\\
\frac{\partial v_y}{\partial a} & \frac{\partial v_y}{\partial e} & \frac{\partial v_y}{\partial i} & \frac{\partial v_y}{\partial \Omega} & \frac{\partial v_y}{\partial \omega} & \frac{\partial v_y}{\partial M_0}\\
\frac{\partial v_z}{\partial a} & \frac{\partial v_z}{\partial e} & \frac{\partial v_z}{\partial i} & \frac{\partial v_z}{\partial \Omega} & \frac{\partial v_z}{\partial \omega} & \frac{\partial v_z}{\partial M_0} 
\end{array} 
\right)\] 
In order to compute these partial derivatives, we consider the relation
between $\vec{r}$ and $\vec{v}$ in a fiducial inertial reference
frame.  These relations can be found in classical mechanical textbooks
such as \citet[][pp. 35-36]{taff1985}.  A standard choice is:

\begin{eqnarray} 
\left(
\begin{array}{c}
r_x 
\\
r_y  
\\
r_z
\end{array}
\right)
=  \cal{R}
\left(
\begin{array}{c}
\frac{a \left(1-e^2\right) \cos{f}}{1+ e\cos{f}}
\\
\frac{a \left(1-e^2\right) \sin{f}}{1+ e\cos{f}}
\\
0
\end{array}
\right),
\qquad
\left(
\begin{array}{c}
v_x 
\\
v_y  
\\
v_z
\end{array}
\right)
=  \mathcal{R}
\left(
\begin{array}{c}
\frac{-n a \sin{f}}{\left(1-e^2\right)^{1/2}}
\\
\frac{n a \left( e+\cos{f} \right)}{\left(1-e^2\right)^{1/2}}
\\ 
0\\
\end{array}
\right)
\end{eqnarray} 
where $\cal{R}$ is the matrix
\begin{equation}
\mathcal{R} =
\left( 
\begin{array}{ccc}
\cos{\Omega} \cos{\omega}\!-\!\sin{\Omega} \sin{\omega} \cos{i} \ & \ -\!\cos{\Omega} \sin{\omega}\!-\!\sin{\Omega} \cos{\omega} \cos{i} \ & \ \sin{\Omega} \sin{i} \\
\sin{\Omega} \cos{\omega}\!+\!\cos{\Omega} \sin{\omega} \cos{i} \ & \ -\!\sin{\Omega} \sin{\omega}\!+\!\cos{\Omega} \cos{\omega} \cos{i} \ & \ -\!\cos{\Omega} \sin{i} \\
\sin{\omega} \sin{i} \ & \ \cos{\omega} \sin{i} \ & \ \cos{i} 
\end{array}
\right) 
\end{equation}
Our choice of fiducial reference frame sets the $x$-axis to be along the
major axis of the ellipse, where the pericenter is in the positive
direction.  The matrices $\mathcal{M}_1, \mathcal{M}_2, \mathcal{M}_3$
are independent of the system being studied.  Their computation does,
however, contain some subtleties which are worth recording: i) the
mean motion $n$ in $\mathcal{M}_3$ is a function of $a$, ii) the true
anomaly $f$ is a function of the time $t$ as well as $M_0$, $e$ {\it
  and} $a$, iii) by definition, $M_0$ is defined at $t=t_0$.
Therefore, we need to compute the partial derivatives with respect to
the true anomaly:

\begin{equation}
\frac{\partial f}{\partial M_0}
=
\frac{\partial f}{\partial E} \frac{\partial E}{\partial M_0}
=
\left(\frac{1 + e \cos{f}}{\sqrt{1-e^2}}\right)
\left(\frac{1 + e \cos{f}}{1-e^2}  \right)
=
\frac{\left(1 + e \cos{f} \right)^2}{\left(1 - e^2\right)^{3/2} } 
\end{equation}

\noindent{where} the expression for the first partial derivative is
from \cite{broucke1970} -- who also provides tables of partial
derivatives, Lagrange Brackets and Poisson Brackets for the
unperturbed two-body problem -- and the second is from Kepler's
equation and from the relations between eccentric anomaly $E$ and the true
anomaly $f$ \footnote{Note that Eq. (111) of \cite{efroimsky2005a} for
  $\partial M/ \partial f$ contains a typo.}:

\begin{equation}
\cos{E} = \frac{e + \cos{f}}
{1 + e \cos{f} },
   \   \   \  \  \
\sin{E} = 
\frac{\sin{f}\sqrt{1 - e^2}}
{1 + e \cos{f}  }
\label{anomid1}
\end{equation}

\noindent{Further}, we have:

\begin{equation}
\frac{\partial f}{\partial e}
=
\frac{\partial f}{\partial E} \frac{\partial E}{\partial e}
=
\left(\frac{1 + e \cos{f}}{\sqrt{1-e^2}}\right)
\left(\frac{a}{r} \sin{E}  \right)
=
\frac{\sin{f}}{1-e^2}\left(2 + e \cos{f}\right)
\end{equation}

\noindent{and}

\begin{equation}
\frac{\partial f}{\partial a}
=
\frac{\partial f}{\partial E} \frac{\partial E}{\partial a}
=
\left(\frac{1 + e \cos{f}}{\sqrt{1-e^2}}\right)
\left(-\frac{3n \left( t-t_0 \right)}{2 r}  \right)
=
-\frac{3}{2} n \left(t-t_0\right) \frac{\left(1 + e \cos{f}\right)^2}{a \left(1-e^2\right)^{3/2}}
\end{equation}

\noindent{where} the partial derivatives of $E$ with respect to $e$
and $a$ are from \cite{broucke1970}.  Importantly, one cannot impose
the condition $t = t_0$ until {\it after} $\mathcal{M}_2$ and
$\mathcal{M}_3$ have been computed.

The time evolution with respect to $E$ or $f$ is often more desirable
to investigators than the time evolution of $M_0$, as is true in this
study.  We can convert one equation into another by taking the total
time derivative of Kepler's equation evaluated at $t = t_0$, where $M
= n(t-t_0) + M_0$.  The result is:
\begin{equation}
\frac{dE}{dt} = \frac{1}{1-e\cos{E}}
\left[    
n + \frac{dM_0}{dt} + \sin{E} \frac{de}{dt} 
\right]
\label{dbigEdt}
\end{equation}
We note that the terms with the time derivatives of the semimajor axis
vanish.  Then we finally obtain:
\begin{equation}
%
\frac{df}{dt} =
\frac{1}{1-e^2}
\left[
\left(1 + e \cos{f} \right) \sqrt{1 - e^2} \frac{dE}{dt} + \sin{f} \frac{de}{dt}
\right]
\label{Ederivf}
\end{equation}

We choose $\vec{\Phi} = 0$ in order to obtain a simply interpreted set of
resulting orbital elements.  However, there may be instances when
choosing a non-zero gauge provides an insight or simplification to the
problem considered.

As a check on these equations, we used them to re-derive the equations
of motion for the two-body problem perturbed by mass loss in
\cite{veretal2011}.

\section{General Equations}

The equations below are well-suited for describing a planet on an
arbitrarily wide and inclined orbit.

\subsection{Orbital Element Equations}

We apply:

\begin{equation}
\Delta \vec{A}
=
\left(
\begin{array}{c}
\Upsilon_{xx} x + \Upsilon_{xy} y + \Upsilon_{uu} \dot{x} + \Upsilon_{uv} \dot{y}
\\
\Upsilon_{yx} x + \Upsilon_{yy} y + \Upsilon_{vu} \dot{x} + \Upsilon_{vv} \dot{y}
\\
\Upsilon_{zz} z + \Upsilon_{ww} \dot{z}
\end{array}
\right)
\label{pre}
\end{equation}
to the formalism in Section 2 in order to derive the equations of
motion in terms of orbital elements:

\begin{eqnarray}
\frac{da}{dt} &=&
  \frac{2a\sqrt{1-e^2}}{n\left(1+e\cos{f}\right)} 
\bigg[ 
\Upsilon_{zz} C_1
       \left\lbrace \sin^2{i} \sin{\left(f+\omega\right)} \right\rbrace
\nonumber
\\
&+&
\left( \Upsilon_{xx} C_3 - \Upsilon_{xy} C_4 \right)
       \left\lbrace C_1 \sin{\Omega} \cos{i} + C_2 \cos{\Omega} \right\rbrace
\nonumber
\\
&-&
\left( \Upsilon_{yx} C_3 - \Upsilon_{yy} C_4 \right)
       \left\lbrace C_1 \cos{\Omega} \cos{i} - C_2 \sin{\Omega} \right\rbrace
\bigg]
\nonumber
\\
&+& \frac{2a}{1 - e^2}
\bigg[
\Upsilon_{ww} \left\lbrace C_{1}^2 \sin^2{i} \right\rbrace
\nonumber
\\
&+&
       \Upsilon_{uu} 
       \left\lbrace C_1 \sin{\Omega} \cos{i} + C_2 \cos{\Omega} \right\rbrace^2
  +    \Upsilon_{vv} 
       \left\lbrace C_1 \cos{\Omega} \cos{i} - C_2 \sin{\Omega} \right\rbrace^2
\nonumber
\\
&-&
       \left(\Upsilon_{uv} + \Upsilon_{vu} \right)
       \left\lbrace C_1 \sin{\Omega} \cos{i} + C_2 \cos{\Omega} \right\rbrace
       \left\lbrace C_1 \cos{\Omega} \cos{i} - C_2 \sin{\Omega} \right\rbrace
\bigg]
\label{gena}
\\
\frac{de}{dt} &=&
  \frac{\left(1-e^2\right)^{\frac{3}{2}}}{2n\left(1+e\cos{f}\right)^2} 
\bigg[ 
\Upsilon_{zz} C_6
       \left\lbrace \sin^2{i} \sin{\left(f+\omega\right)} \right\rbrace
\nonumber
\\
&+&
\left( \Upsilon_{xx} C_3 - \Upsilon_{xy} C_4 \right)
       \left\lbrace C_6 \sin{\Omega} \cos{i} + C_5 \cos{\Omega} \right\rbrace
\nonumber
\\
&-&
\left(\Upsilon_{yx} C_3 - \Upsilon_{yy} C_4 \right)
       \left\lbrace C_6 \cos{\Omega} \cos{i} - C_5 \sin{\Omega} \right\rbrace
\bigg]
\nonumber
\\
&+& \frac{1}{2 \left(1 + e \cos{f}\right)}
\bigg[
\Upsilon_{ww} \left\lbrace C_{1} C_6 \sin^2{i} \right\rbrace
\nonumber
\\
&+&
       \Upsilon_{uu} 
       \left\lbrace C_1 \sin{\Omega} \cos{i} + C_2 \cos{\Omega} \right\rbrace
       \left\lbrace C_6 \sin{\Omega} \cos{i} + C_5 \cos{\Omega} \right\rbrace
\nonumber
\\
&+&    \Upsilon_{vv} 
       \left\lbrace C_1 \cos{\Omega} \cos{i} - C_2 \sin{\Omega} \right\rbrace
       \left\lbrace C_6 \cos{\Omega} \cos{i} - C_5 \sin{\Omega} \right\rbrace
\nonumber
\\
&-&    \Upsilon_{uv}
       \left\lbrace C_1 \cos{\Omega} \cos{i} - C_2 \sin{\Omega} \right\rbrace
       \left\lbrace C_6 \sin{\Omega} \cos{i} + C_5 \cos{\Omega} \right\rbrace
\nonumber
\\
&+&    \Upsilon_{vu}
       \left\lbrace C_1 \sin{\Omega} \cos{i} + C_2 \cos{\Omega} \right\rbrace
       \left\lbrace -C_6 \cos{\Omega} \cos{i} + C_5 \sin{\Omega} \right\rbrace
\bigg]
\label{gene}
\\
\frac{di}{dt} &=& 
  \frac{\left(1-e^2\right)^{\frac{3}{2}} \sin{i}}{n\left(1+e\cos{f}\right)^2} 
\bigg[ 
\Upsilon_{zz} 
       \left\lbrace \cos{i} \cos{\left(f+\omega\right)} \sin{\left(f+\omega\right)} \right\rbrace
\nonumber
\\
&-&
\left( \Upsilon_{xx} C_3 - \Upsilon_{xy} C_4 \right)
       \left\lbrace \sin{\Omega} \cos{\left(f+\omega\right)} \right\rbrace
+
\left( \Upsilon_{yx} C_3 - \Upsilon_{yy} C_4 \right)
       \left\lbrace \cos{\Omega} \cos{\left(f+\omega\right)}\right\rbrace
\bigg]
\nonumber
\\
&+& \frac{\cos{\left(f+\omega\right)} \sin{i}}{1 + e \cos{f}}
\bigg[
\Upsilon_{ww} \left\lbrace C_{1} \cos{i} \right\rbrace
\nonumber
\\
&-&
       \Upsilon_{uu} \sin{\Omega} 
       \left\lbrace C_1 \sin{\Omega} \cos{i} + C_2 \cos{\Omega} \right\rbrace
+    
       \Upsilon_{vv} \cos{\Omega}
       \left\lbrace -C_1 \cos{\Omega} \cos{i} + C_2 \sin{\Omega} \right\rbrace
\nonumber
\\
&+&    \Upsilon_{uv} \sin{\Omega}
       \left\lbrace C_1 \cos{\Omega} \cos{i} - C_2 \sin{\Omega} \right\rbrace
+    
       \Upsilon_{vu} \cos{\Omega}
       \left\lbrace C_1 \sin{\Omega} \cos{i} + C_2 \cos{\Omega} \right\rbrace
\bigg]
\label{geni}
\\
\frac{d\Omega}{dt} &=& 
  \frac{\left(1-e^2\right)^{\frac{3}{2}}}{n\left(1+e\cos{f}\right)^2} 
\bigg[ 
\Upsilon_{zz} 
       \left\lbrace \cos{i} \sin^2{\left(f+\omega\right)} \right\rbrace
\nonumber
\\
&-&
\left( \Upsilon_{xx} C_3 - \Upsilon_{xy} C_4 \right)
       \left\lbrace \sin{\Omega} \sin{\left(f+\omega\right)} \right\rbrace
+
\left( \Upsilon_{yx} C_3 - \Upsilon_{yy} C_4 \right)
       \left\lbrace \cos{\Omega} \sin{\left(f+\omega\right)}\right\rbrace
\bigg]
\nonumber
\\
&+& \frac{\sin{\left(f+\omega\right)}}{1 + e \cos{f}}
\bigg[
\Upsilon_{ww} \left\lbrace C_{1} \cos{i} \right\rbrace
\nonumber
\\
&-&
       \Upsilon_{uu} \sin{\Omega} 
       \left\lbrace C_1 \sin{\Omega} \cos{i} + C_2 \cos{\Omega} \right\rbrace
+    
       \Upsilon_{vv} \cos{\Omega}
       \left\lbrace -C_1 \cos{\Omega} \cos{i} + C_2 \sin{\Omega} \right\rbrace
\nonumber
\\
&+&    \Upsilon_{uv} \sin{\Omega}
       \left\lbrace C_1 \cos{\Omega} \cos{i} - C_2 \sin{\Omega} \right\rbrace
+    
       \Upsilon_{vu} \cos{\Omega}
       \left\lbrace C_1 \sin{\Omega} \cos{i} + C_2 \cos{\Omega} \right\rbrace
\bigg]
\label{genO}
\\
\frac{d\omega}{dt} &=&     
  \frac{\left(1-e^2\right)^{\frac{3}{2}}}{2en\left(1+e\cos{f}\right)^2} 
\bigg[ 
\Upsilon_{zz} 
       \left\lbrace -2\sin{\left(f+\omega\right)}
  \left[ e \sin{\left(f+\omega\right)} + \frac{1}{2} C_8 \sin^2{i} \right]
       \right\rbrace
\nonumber
\\
&-&
\left( \Upsilon_{xx} C_3 - \Upsilon_{xy} C_4 \right)
       \left\lbrace C_8 \sin{\Omega} \cos{i} - C_7 \cos{\Omega} \right\rbrace
\nonumber
\\
&+&
\left( \Upsilon_{yx} C_3 - \Upsilon_{yy} C_4 \right)
       \left\lbrace C_8 \cos{\Omega} \cos{i} + C_7 \sin{\Omega} \right\rbrace
\bigg]
\nonumber
\\
&+& \frac{1}{2 e \left(1 + e \cos{f}\right)}
\bigg[
C_1 \Upsilon_{ww} \big\lbrace 
                 -2 e \left( \cos{f} \sin{\omega} + \cos{\omega} \sin{f} \cos^2{i} \right)
\nonumber
\\                 
&+&  
\sin^2{i} \left(\sin{\left(2f + \omega\right)} - 3 \sin{\omega} \right)   
                 \big\rbrace
\nonumber
\\
&+&
       \Upsilon_{uu} 
       \left\lbrace C_1 \sin{\Omega} \cos{i} + C_2 \cos{\Omega} \right\rbrace
       \left\lbrace -C_8 \sin{\Omega} \cos{i} + C_7 \cos{\Omega} \right\rbrace
\nonumber
\\
&-&    \Upsilon_{vv} 
       \left\lbrace C_1 \cos{\Omega} \cos{i} - C_2 \sin{\Omega} \right\rbrace
       \left\lbrace C_8 \cos{\Omega} \cos{i} + C_7 \sin{\Omega} \right\rbrace
\nonumber
\\
&-&    \Upsilon_{uv}
       \left\lbrace C_1 \cos{\Omega} \cos{i} - C_2 \sin{\Omega} \right\rbrace
       \left\lbrace -C_8 \sin{\Omega} \cos{i} + C_7 \cos{\Omega} \right\rbrace
\nonumber
\\
&+&    \Upsilon_{vu}
       \left\lbrace C_1 \sin{\Omega} \cos{i} + C_2 \cos{\Omega} \right\rbrace
       \left\lbrace C_8 \cos{\Omega} \cos{i} + C_7 \sin{\Omega} \right\rbrace
\bigg]
\label{geno}
        \\
\frac{df}{dt}   &=& \frac{ n
\left( 1 + e \cos{f} \right)^2 }
{\left(1 - e^2 \right)^{3/2}}
+
  \frac{\left(1-e^2\right)^{\frac{3}{2}}}{2en\left(1+e\cos{f}\right)^2} 
\bigg[ 
\Upsilon_{zz} 
       \left\lbrace \sin^2{i} \sin{\left(f+\omega\right) 
             \left[ C_8 + 2 e \sin{\left(f+\omega\right)} \right]
     } \right\rbrace
\nonumber
\\
&+&
\left( \Upsilon_{xx} C_3 - \Upsilon_{xy} C_4 \right)
       \left\lbrace C_9 \sin{\Omega}\cos{i} - C_7 \cos{\Omega} \right\rbrace
\nonumber
\\
&-&
\left( \Upsilon_{yx} C_3 - \Upsilon_{yy} C_4 \right)
       \left\lbrace C_9 \cos{\Omega}\cos{i}  + C_7 \sin{\Omega}   \right\rbrace
\bigg]
\nonumber
\\
&+& \frac{1}{2 e \left(1 + e \cos{f}\right)}
\bigg[
    \Upsilon_{ww} \left\lbrace C_1 C_9 \sin^2{i} \right\rbrace
\nonumber
\\
&-&
       \Upsilon_{uu} 
       \left\lbrace C_1 \sin{\Omega} \cos{i} + C_2 \cos{\Omega} \right\rbrace
       \left\lbrace -C_9 \sin{\Omega} \cos{i} + C_7 \cos{\Omega} \right\rbrace
\nonumber
\\
&+&    \Upsilon_{vv} 
       \left\lbrace C_1 \cos{\Omega} \cos{i} - C_2 \sin{\Omega} \right\rbrace
       \left\lbrace C_9 \cos{\Omega} \cos{i} + C_7 \sin{\Omega} \right\rbrace
\nonumber
\\
&+&    \Upsilon_{uv}
       \left\lbrace C_1 \cos{\Omega} \cos{i} - C_2 \sin{\Omega} \right\rbrace
       \left\lbrace -C_9 \sin{\Omega} \cos{i} + C_7 \cos{\Omega} \right\rbrace
\nonumber
\\
&-&    \Upsilon_{vu}
       \left\lbrace C_1 \sin{\Omega} \cos{i} + C_2 \cos{\Omega} \right\rbrace
       \left\lbrace C_9 \cos{\Omega} \cos{i} + C_7 \sin{\Omega} \right\rbrace
\bigg]
\label{genf1}
\\
&=&
-\frac{d\omega}{dt} - \cos{i} \frac{d\Omega}{dt}
+
\frac{ n
\left( 1 + e \cos{f} \right)^2 }
{\left(1 - e^2 \right)^{3/2}}
\label{genf}
\end{eqnarray}
where the quantities $C_1, \dots C_9$ are given in Appendix A.  All $C$
terms are on the order of unity, and so the only variables in the
square brackets not of the order of unity are the $\Upsilon$ values.

Note the remarkably simple form taken on by $df/dt$ in
Eq. (\ref{genf}), where the last term represents the two-body term.
This form allows us to quickly derive some other results.  The
argument of latitude $\upsilon = f + \omega$ is a
quantity which evolves with time according to:
\begin{equation}
\frac{d\upsilon}{dt}  = - \cos{i} \frac{d\Omega}{dt}
+
\frac{ n
\left( 1 + e \cos{f} \right)^2 }
{\left(1 - e^2 \right)^{3/2}}
\label{arglat}
\end{equation}
The evolution of the true longitude $\theta = \upsilon + \Omega$ is:
\begin{equation}
\frac{d\theta}{dt} = \left(1 - \cos{i}\right) \frac{d\Omega}{dt}
+
\frac{ n
\left( 1 + e \cos{f} \right)^2 }
{\left(1 - e^2 \right)^{3/2}}
\label{trulon}
\end{equation}

\subsection{Properties} \label{firprop}

Equations (\ref{gena}) - (\ref{genf}) are completely equivalent to
Eqs. (\ref{xeq}) - (\ref{zeq}).  However, Eqs. (\ref{gena}) -
(\ref{genf}) let us deduce properties of planetary motion, which may
not otherwise be apparent:

First, note that all the $\Upsilon$ terms are decoupled, allowing us
to easily consider special-case physical situations by striking out
the relevant terms. The rates of change of eccentricity, inclination,
longitude of ascending node and longitude of pericenter ($|de/dt|$,
$|di/dt|$, $|d\Omega/dt|$ and $|d\omega/dt|$) are all $\propto 1/n
\propto a^{3/2}$ for the position-only $\Upsilon$ terms, whereas the
rate of change of semimajor axis $|da/dt| \propto a/n \propto
a^{5/2}$. Turning to velocity-based $\Upsilon$ terms, $|de/dt|$,
$|di/dt|$, $|d\Omega/dt|$ and $|d\omega/dt|$ are all independent of
$a$, whereas $|da/dt| \propto a$.

We can also deduce some properties of the evolution of special
orbits.  Planar orbits ($i=0$) {\it will} remain planar, but circular
orbits {\it will not} remain circular. Likewise, polar orbits {\it
  will not} remain polar.  Further, the sign of $di/dt$ will be
different depending on whether $i=90^{\circ}$ or $i=270^{\circ}$.

As $|d\omega/dt| \propto 1/e$ and $|df/dt| \propto 1/e$, then for
circular orbits, neither $\omega$ nor $f$ are well-defined physically
nor mathematically.  Although $\Omega$ is not physically defined for
planar orbits, $\Omega$ still remains well-defined mathematically, and
remains an independent variable in all the evolution equations.  In
order to use a physically meaningful variable in the planar case, one
must convert all of the equations to a different variable, such as
$\varpi \equiv \Omega + \omega$.

Equation (\ref{trulon}) allows us to assess how the planet appears to move
from the point of view of a fixed observer: a planet will still appear
to circulate around its parent star without ever reversing direction
on the sky subject to all these external forces only if the planet is
on a planar orbit.  The extent of any modulation in the apparent
motion, provided entirely by $d\Omega/dt$, increases as the planet's
orbit becomes increasingly non-coplanar.

\section{Adiabatic Limit}

A planet on an orbit which is compact enough so that its period is
smaller than the timescale of any external perturbations can be treated in the adiabatic
approximation.  This approximation has been utilized in many facets of
planetary astronomy because the majority of exoplanets discovered are
within hundreds of AU of their parent stars.

\subsection{Orbital Element Equations}

For the remainder of this paper, in order to simplify and focus our
results, we assume $\Upsilon_{uu} = \Upsilon_{vv} = \Upsilon_{ww} =
0$.  These terms do not play a role in many applications, such as
tidal distortions of planetary orbits  \citep{braetal2010}, but were
included in the general equations for completeness and future studies.

Here, we consider the ``adiabatic''  case, where averaging over
the fast variables is justified. This corresponds to the following
inequalities being satisfied: $\Upsilon_{xx}/n^2 \ll 1$, 
$\Upsilon_{xy}/n^2 \ll 1$, $\Upsilon_{yx}/n^2 \ll 1$,
$\Upsilon_{yy}/n^2 \ll 1$, $\Upsilon_{zz}/n^2 \ll 1$,
$\Upsilon_{uv}/n \ll 1$, $\Upsilon_{vu}/n \ll 1$.  We refer to 
this approximation as adiabatic, and this condition renders the equations of motion
independent of the true anomaly $f$.  We assume

\begin{equation}
\frac{dt}{df} = \frac{\left(1-e^2\right)^{3/2}}{n \left(1 + e \cos{f}\right)^2},
\end{equation}
which is the unperturbed two-body term.  We obtain the adiabatic limit
by computing:
\begin{equation}
\frac{d\beta}{dt}_{\rm adiabatic} = 
\frac{n}{2\pi}\int_{0}^{2\pi} \frac{d\beta}{dt}_{\rm non-adiabatic} \frac{dt}{df} df
\end{equation}
for a general variable $\beta$.  Carrying out the averaging, we find:
\begin{eqnarray}
\frac{da}{dt}
&=&
\frac{a \sqrt{1 - e^2} \cos{i}}{n} 
\left(\Upsilon_{yx} - \Upsilon_{xy} \right)
\nonumber
\\
&-&
\frac{a}{8e^2}
\bigg[ \left(-8 + 4e^2 + 8 \sqrt{1-e^2}  \right) \cos{(2\Omega)} \sin{(2\omega)} \cos{i}
\nonumber
\\
&+&
\bigg(
\left( -2 + e^2 + 2 \sqrt{1-e^2} \right)
\left( 3 +  \cos{2i} \right)
\cos{2\omega}
\nonumber
\\
&-&
2e^2\sin^2{i}
\bigg)
\sin{2\Omega}
   \bigg]
\left(\Upsilon_{uv} + \Upsilon_{vu} \right)
\label{adia}
\\
\frac{de}{dt}
&=&
-\frac{5e \sqrt{1 - e^2}}{2 n} \cos{\omega} \sin{\omega} \sin^2{i}
\Upsilon_{zz} 
+
\frac{5e \sqrt{1 - e^2}}{16 n} \bigg\lbrace
\nonumber
\\
&+&
\left[
4 \cos{i} \cos{2\omega} \sin{2\Omega} 
+
\sin{2\omega}
\left(  
\cos{2\Omega} \left(3 + \cos{2i} \right) + 2 \sin^2{i}
\right)
\right]
\Upsilon_{xx}
 \nonumber
\\
&-&
\left[
4 \cos{i} \cos{2\omega} \sin{2\Omega}
+
\sin{2\omega}
\left(  
\cos{2\Omega} \left(3 + \cos{2i} \right) - 2 \sin^2{i}
\right)
\right]
\Upsilon_{yy} 
\nonumber
\\
&-&
\left[
4 \cos{i}\cos{2\omega} \cos{2\Omega} - 4 \cos{i}
-
\sin{2\omega}\sin{2\Omega} \left(3 + \cos{2i} \right)
\right]
\Upsilon_{xy}
\nonumber
\\
&-&
\left[
4 \cos{i}\cos{2\omega} \cos{2\Omega} + 4 \cos{i}
-
\sin{2\omega}\sin{2\Omega} \left(3 + \cos{2i} \right)
\right]
\Upsilon_{yx}
\bigg\rbrace
\nonumber
\\
&+&
\frac{1}{8e^3} 
\bigg[
\left(1 - e^2\right)
\left(2 - e^2 - 2\sqrt{1-e^2}\right)
\nonumber
\\
&\times& \left(4 \cos{i} \cos{2\Omega} \sin{2\omega} + 
\left(3 + \cos{2i} \right)
\cos{2\omega}
\sin{2\Omega}
\right)
\bigg]
\left(\Upsilon_{uv}+ \Upsilon_{vu}\right)
\label{adie}
\\
\frac{di}{dt}
&=&  
\frac{5e^2\sin{2\omega}\sin{2i}}{8n\sqrt{1 - e^2}}
\Upsilon_{zz} 
+ \frac{\sin{i}}{8n\sqrt{1 - e^2}}
\bigg\lbrace
\nonumber
\\
&+&
\left[ \sin{2\Omega} \left(2 + 3 e^2 + 5 e^2 \cos{2 \omega} \right)
-
10 e^2 \cos{i} \sin{2\omega} \sin^2{\Omega}  \right]
\Upsilon_{xx}
\nonumber
\\
&-&
\left[ \sin{2\Omega} \left(2 + 3 e^2 + 5 e^2 \cos{2 \omega} \right)
+
10 e^2 \cos{i} \sin{2\omega} \cos^2{\Omega}  \right]
\Upsilon_{yy}
\nonumber
\\
&+&
\left[
2 \sin^2{\Omega}  \left(2 + 3 e^2 + 5 e^2 \cos{2 \omega} \right)
+
10 e^2 \cos{i} \sin{2\omega} \cos{\Omega} \sin{\Omega}
\right]
\Upsilon_{xy}
\nonumber
\\
&-&
\left[
2 \cos^2{\Omega}  \left(2 + 3 e^2 + 5 e^2 \cos{2 \omega} \right)
-
10 e^2 \cos{i} \sin{2\omega} \cos{\Omega} \sin{\Omega}
\right]
\Upsilon_{yx} \bigg\rbrace
\nonumber
\\
&+&
\frac{1}{8} \sin{2i}\sin{2\Omega} \left(\Upsilon_{uv} + \Upsilon_{vu}\right)
\label{adii}
\\
\frac{d\Omega}{dt}
&=& 
\frac{\cos{i} \left(2 + 3 e^2 - 5 e^2 \cos{2 \omega} \right)}{4n\sqrt{1 - e^2}}
\Upsilon_{zz}
\nonumber
+
\frac{1}{4n\sqrt{1 - e^2}} \bigg\lbrace
\\
&+&
\left[ \sin^2{\Omega} \cos{i} \left(-2 - 3 e^2 + 5 e^2 \cos{2 \omega} \right)
+
5 e^2 \cos{\Omega} \sin{\Omega} \sin{2\omega}  \right]
\Upsilon_{xx}
\nonumber
\\
&+& 
\left[ \cos^2{\Omega} \cos{i} \left(-2 - 3 e^2 + 5 e^2 \cos{2 \omega} \right)
-
5 e^2 \cos{\Omega} \sin{\Omega} \sin{2\omega}  \right]
\Upsilon_{yy}
\nonumber
\\
&-& 
\left[ \cos{\Omega} \sin{\Omega} \cos{i} \left(-2 - 3 e^2 + 5 e^2 \cos{2 \omega} \right)
-
5 e^2 \sin^2{\Omega} \sin{2\omega}  \right]
\Upsilon_{xy}
\nonumber
\\
&-&
\left[ \cos{\Omega} \sin{\Omega} \cos{i} \left(-2 - 3 e^2 + 5 e^2 \cos{2 \omega} \right)
+
5 e^2 \cos^2{\Omega} \sin{2\omega}  \right]
\Upsilon_{yx} \bigg\rbrace
\nonumber
\\
&-&
\left(\frac{1}{2} \sin^2{\Omega}\right) \Upsilon_{uv}
+
\left(\frac{1}{2} \cos^2{\Omega}\right) \Upsilon_{vu}
\label{adOi}
\\
\frac{d\omega}{dt}
&=&
\frac{5\sin^2{\omega} \left(\sin^2i - e^2\right) - \left(1 - e^2\right)}{2n\sqrt{1-e^2}}
\Upsilon_{zz}
+ \frac{1}{16n\sqrt{1-e^2}} \bigg\lbrace
\nonumber
\\
&+&
\left( 2C_{10} + C_{11} + C_{12} \right) \Upsilon_{xx}
+
\left( -2C_{10} - C_{11} + C_{12} \right) \Upsilon_{yy}
\nonumber
\\
&+&
\left( -2C_{10}\cot{2\Omega} - C_{13} - 10 e^2 \cos{i} \sin{2\omega} \right) \Upsilon_{xy}
\nonumber
\\
&+&
\left( -2C_{10}\cot{2\Omega} - C_{13} + 10 e^2 \cos{i} \sin{2\omega} \right) \Upsilon_{yx}
\bigg\rbrace
\nonumber
\\
&+&
\frac{1}{16e^4} \left(-4C_{14} + C_{15}\right) \left( \Upsilon_{uv} + \Upsilon_{vu} \right)
\label{adivarpi}
\end{eqnarray}
where the variables $C_{10} \dots C_{15}$ are given in Appendix A.
When only vertical external forces are included, the equations yield
the same stationary solution as in \cite{brasser2001}.  In the more
general case, when the non-vertical external forces are included, the
equations reduce to Eqs. (3)-(7) of \cite{fouchard2004} and the
equations in Appendix A of \cite{fouetal2006}, given their assumptions
about $\Upsilon$ and after converting their Delaunay elements to the
orbital elements used in this work.

The $\Upsilon_{zz}, \Upsilon_{xx}$ and $\Upsilon_{yy}$ terms in the expression for
$da/dt$ vanish.  These terms are non-zero in all other orbital element
equations.  Therefore, a planet's semimajor axis is never secularly
affected by vertical perturbations.  In no case does orbit-averaging
cause a velocity-dependent $\Upsilon$ term to vanish.  However, in all
cases, except for $d\Omega/dt$, both velocity dependent terms appear
in the combination $(\Upsilon_{uv} + \Upsilon_{vu})$, which does
vanish when there is no net shear on the two-body system.

\section{Planar Adiabatic Limit} \label{sec5planar}

In the planar adiabatic limit, the general equations are greatly
simplified.  If the perturbations allow the planet to remain in its
original orbital plane, then the vertical contributions vanish.  The
planet may have a planar orbit in two ways: in a prograde ($i =
0^{\circ}$) or retrograde sense ($i = 180^{\circ}$).  As mentioned
previously, because $\Omega$ is not defined physically for
$i=0^{\circ}$ or $i=180^{\circ}$, we wish to eliminate it from the
equations of motion in the planar case\footnote{Mathematically, in the
  planar case, $\Omega \ne 0^{\circ}$ and changes with time, as can be
  seen in Eq. (\ref{genO}).}.  Trigonometric manipulation demonstrates
that in both the prograde and retrograde cases, both $\Omega$ and
$\omega$ will vanish when a suitable angle is defined.

\subsection{Prograde Equations}

In the prograde case, we use the longitude of pericenter $\varpi =
\Omega + \omega$ and $i = 0^{\circ}$. When the planet's orbit
is sufficiently small, then we can impose the adiabatic approximation to obtain
\begin{eqnarray}
\frac{da}{dt}
&=& -\frac{a \sqrt{1-e^2}}{n} \left(  \Upsilon_{xy} - \Upsilon_{yx}  \right)
- \frac{a \left(e^2 - 2 + 2\sqrt{1-e^2}\right) \sin{2\varpi}}
{2 e^2} 
\left(  \Upsilon_{uv} + \Upsilon_{vu}  \right)
\label{ppaa}
     \\
\frac{de}{dt}
&=& 
\frac{5e\sqrt{1-e^2}}{4n}
\left[ \sin{2\varpi} \left(\Upsilon_{xx} - \Upsilon_{yy} \right) + 
\left(2\sin^2{\varpi} \right)\Upsilon_{xy} - \left(2\cos^2{\varpi} \right)\Upsilon_{yx} \right]
\nonumber
\\
&-&
\frac{\left(1 - e^2\right) \left(2 - e^2 - 2\sqrt{1 - e^2} \right) \sin{2\varpi}}{2e^3}
\left(  \Upsilon_{uv} + \Upsilon_{vu}  \right)
\label{ppae}
    \\
\frac{di}{dt}
&=& 0   
     \\
\frac{d\varpi}{dt}
&=&   
\frac{\sqrt{1-e^2}}{4n} 
\left[  
\left(3 + 5 \cos{2\varpi} \right) \Upsilon_{xx}
+ \left(3 - 5 \cos{2\varpi} \right) \Upsilon_{yy}
+ 5 \sin{2\varpi} \left( \Upsilon_{xy} + \Upsilon_{yx}  \right)
\right]
\nonumber
\\
&-&
\bigg[
\frac{1}{4} 
\left( \Upsilon_{uv} - \Upsilon_{vu} \right) 
\nonumber
\\
&+&
\frac{\left(4 + e^4 - 6e^2 - 4 \left(1 - e^2\right)^{3/2}\right) \cos{2\varpi}}
{4e^4}
\left( \Upsilon_{uv} + \Upsilon_{vu} \right) 
\bigg]
\label{ppaom}
\end{eqnarray}
If $(\Upsilon_{uv} + \Upsilon_{vu}) \ne 0$, then the effect on the
orbit may be drastic, even for orbiting ``Hot Jupiters'' at $a \approx 0.05$ AU.  
The terms which represent the coefficients of $(\Upsilon_{uv} +
\Upsilon_{vu})$ for $da/dt$, $de/dt$ and $d\varpi/dt$ are well-defined
in the limits of $e \rightarrow 0$ and $e \rightarrow 1$.  Further,
the term for $de/dt$ takes on a maximum of $\approx 0.06$ at $e =
\sqrt{2\sqrt{3} - 3} \approx 0.68$.  This term is independent of $a$
and $n$ and hence might dominate the eccentricity evolution.

Note too that the pericenter will be constantly perturbed by the
external force regardless of any properties of the planet unless
$\Upsilon_{uv} = \Upsilon_{vu} = 0$; even if $\Upsilon_{uv} =
\Upsilon_{vu} \ne 0$ and $e = 0$, the last term in Eq. (\ref{ppaom})
will immediately become non-zero.

\subsection{Retrograde Equations}

The equations of motion for the completely retrograde case ($i =
180^{\circ}$) may take on a similar form if a different angle, an {\it
  obverse of pericenter}, $\varsigma$, is defined to be the dog-leg
angle equal to the sum of the longitude of descending node and the
angle between the radius vector of the descending node and the
pericenter of the orbit ($\varsigma \equiv 180^{\circ} - \Omega +
180^{\circ} + \omega = \omega - \Omega$).  Defining this angle allows
us to eliminate both $\Omega$ and $\omega$, as in the previous
subsection.  Both nodes may be physically important; the longstanding
prevalence of the ascending node in dynamical parlance is perhaps due
to the abundance of prograde orbits in the Solar System and, likely,
exoplanetary systems, where prograde is defined with respect to the
parent star's rotation.

Thus, for the planar retrograde case, we find
\begin{eqnarray}
\frac{da}{dt}
&=& \frac{a \sqrt{1-e^2}}{n} \left(  \Upsilon_{xy} - \Upsilon_{yx}  \right)
+ \frac{a \left(e^2 - 2 + 2\sqrt{1-e^2}\right) \sin{2\varsigma}}
{2 e^2} 
\left(  \Upsilon_{uv} + \Upsilon_{vu}  \right)
\label{praa}
     \\
\frac{de}{dt}
&=& 
\frac{5e\sqrt{1-e^2}}{4n}
\left[ \sin{2\varsigma} \left( \Upsilon_{xx} - \Upsilon_{yy} \right)
 - \left( 2\sin^2{\varsigma} \right)\Upsilon_{xy} + \left( 2\cos^2{\varsigma} \right)\Upsilon_{yx} \right]
\nonumber
\\
&+&
\frac{\left(1 - e^2\right) \left(2 - e^2 - 2\sqrt{1 - e^2} \right) \sin{2\varsigma}}{2e^3}
\left(  \Upsilon_{uv} + \Upsilon_{vu}  \right)
\label{prae}
    \\
\frac{di}{dt}
&=& 0   
     \\
\frac{d\varsigma}{dt}
&=&   
\frac{\sqrt{1-e^2}}{4n} 
\left[  
\left(3 + 5 \cos{2\varsigma} \right) \Upsilon_{xx}
+ \left(3 - 5 \cos{2\varsigma} \right) \Upsilon_{yy}
- 5 \sin{2\varsigma} \left( \Upsilon_{xy} + \Upsilon_{yx} \right) 
\right]
\nonumber
\\
&+&
\bigg[
\frac{1}{4} 
\left( \Upsilon_{uv} - \Upsilon_{vu} \right) 
\nonumber
\\
&+&
\frac{\left(4 + e^4 - 6e^2 - 4 \left(1 - e^2\right)^{3/2}\right)\cos{2\varsigma}}
{4e^4}
\left( \Upsilon_{uv} + \Upsilon_{vu} \right) 
\bigg]
\label{praom}
\end{eqnarray}
Note that there is at least one sign change in a term from the
prograde case for each of the expressions for $da/dt$, $de/dt$ and
$d\varsigma/dt$.

\section{An Application: Galactic Tides}

One potential application of the equations is the modelling of
Galactic tides.  There has been previous work on the effect of the
Galactic tide on the Oort Cloud, as this tide strongly affects
the orbital evolution of cloud comets~\citep{heitre1986,matwhi1989}.  At the
Sun's location in the Galaxy, the dominant component of the tide is
caused by the Galactic disk and it acts in a direction perpendicular
to the disk. For the planets in the Solar system, the Galactic tide is
not generally important.  A rough rule-of-thumb is that the precession
due to tides $P_{\rm tide} \approx P_{\rm ext}^2/P_{\rm pl}$, where
$P_{\rm ext}$ is the orbital period of the host star in the Galaxy and
$P_{\rm pl}$ is the orbital period of the plant around the star. For
Jupiter, this gives $P_{\rm tide} \approx 10^{14}$ yr, well in excess
of a Hubble time. The timescale for the effects of the Galactic tide
on the planets in the Solar system to manifest themselves is very
long.

However, the growing consensus that exoplanets are ubiquitous
throughout the Galaxy, together with the discovery of wide-orbit
($\approx$1000-2500 AU) planets \citep[e.g.][]{kuzetal2011,luhetal2011}, 
motivates the study
of the effects of Galactic tides much more generally than has been
done hitherto.

If the host star moving on a circular orbit in the Galaxy, then the
equations for the effects of Galactic tides on the satellite of a are
given by \cite{heitre1986} and \cite{braetal2010}.  In this picture,
the frame of reference is centered on the star, orbiting with the star
but not rotating.  The frame is non-inertial.  Assuming a completely
flat Galactic rotation curve, the only non-vanishing planar terms are
\begin{eqnarray}
\Upsilon_{xx} =  \OmegaG^2 \cos{\left( 2 \OmegaG t \right)}
\quad\quad & &
\Upsilon_{xy} =  \OmegaG^2 \sin{\left( 2 \OmegaG t \right)} \nonumber
\\
\Upsilon_{yx} =  \OmegaG^2 \sin{\left( 2 \OmegaG t \right)}
\quad\quad & &
\Upsilon_{yy} =  -\OmegaG^2 \cos{\left( 2 \OmegaG t \right)}
 \end{eqnarray}
where $\OmegaG$ is the circular frequency of the star with respect to
the Galactic center.  The only non-vanishing vertical term is
$\Upsilon_{zz} = -4 \pi G \rhoG$, where $\rhoG$ is the local Galactic
matter density (including both the luminous matter and any contribution from
dark matter)\footnote{There is a sign error in the corresponding 
equation, the third Eq. 3, in \cite{braetal2010}.}.  
As characteristic of the Galactocentric distances
probed by microlensing -- exemplified by OGLE
2007-BLG-050~\citep{batista2009} and
OGLE-2003-BLG-235~\citep{bennett2006} -- we take the location of the
host star to be 3 kpc from the Galactic Center. 
We make the simple assumption that the Galactic
rotation curve is flat with amplitude $220$ kms$^{-1}$.  Thus,
for a Galactocentric distance $R$, we have
$\OmegaG = 220$ kms$^{-1}/R$.  Further, at $R = 3$ kpc, we set 
$\rhoG = 0.65 M_{\odot}$pc$^{-3}$, self-consistently generating the rotation curve
(see e.g., Eq. 2.2 of \citealt*{evans1993}).

At a distance of 3 kpc, wide-orbit planets may no longer be bound
to their parent star.  One way to quantify this boundary is to compute
the Hill radius of the star, $R_H$.  Equation (20) of \cite{jiatre2010} suggests 
that $R_H = [Gm_{\star}/(2 \Omega_{G}^2)]^{1/3}$.  Using $\Omega_G = 220 {\rm kms}^{-1}/R$, 
we find $R_H = 7.3 \times 10^4 {\rm AU} (R/{\rm kpc})^{2/3}$.  Therefore, for $R=3$ kpc,
the Hill radius is at least $2 \times 10^5$ AU, suggesting that any orbiting
objects within that distance will remain bound.

We now apply these tides to the wide-orbit planet WD 0806-661B b, Sedna-like objects,
and distant scattered disk objects.  This last example provides a demonstration of when the adiabatic
approximation breaks down, and hence when Eqs. (\ref{gena})-(\ref{genf}) must be used instead of
Eqs. (\ref{adia})-(\ref{adivarpi}).  For every curve on every plot, we integrated the equations of 
motion in both orbital elements and Cartesian elements to check that the results are equivalent.

\begin{figure}
\centerline{\put(50,0){\LARGE {{\bf \ \ \ \ \ \ \ \ \ \ \ Widest-Orbit Exoplanet}}}}
\centerline{
  \includegraphics[width=0.70\textwidth,height=6cm]{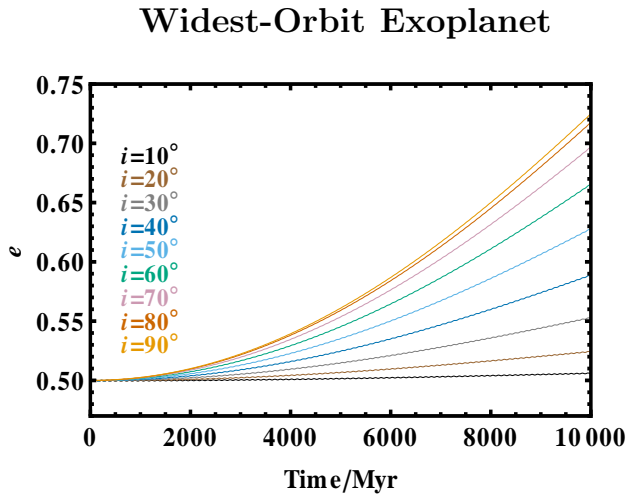}
}
\caption{The eccentricity evolution of the exoplanet with the widest-known orbit,
WD 0806-661B b, due to Galactic tides.  Assumed initial parameters are
$a_0 = 2500$ AU, $e_0 = 0.5, \omega_0 = 0^{\circ}, \Omega_0 = 0^{\circ}$.}
\label{FigWideOrbit}   
\end{figure}


\begin{figure*}
\centerline{\put(50,0){\LARGE {{\bf \ \ \ \ \ \ \ \ \ \ \ \ \ \ \ \ Sedna-like Bodies}}}}
\centerline{
  \includegraphics[width=0.70\textwidth,height=6cm]{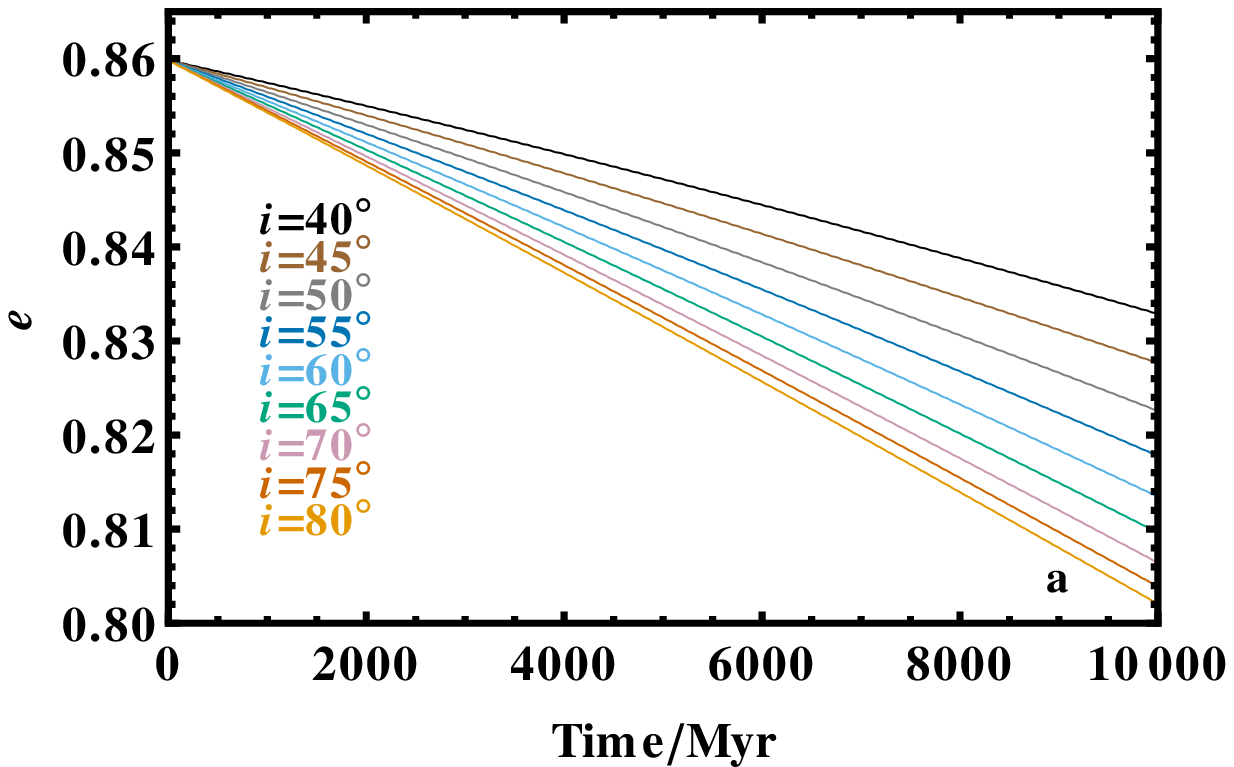}
}
\centerline{
  \includegraphics[width=0.70\textwidth,height=6cm]{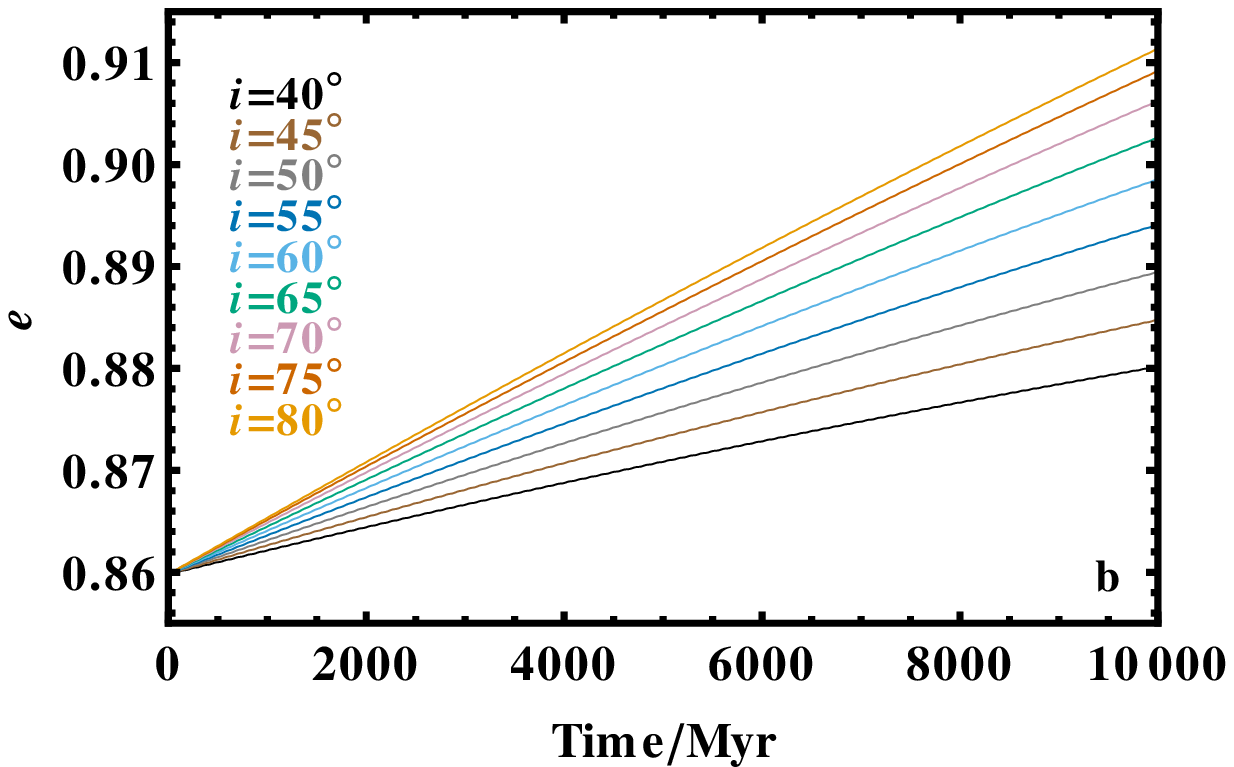}
}
\centerline{
  \includegraphics[width=0.70\textwidth,height=6cm]{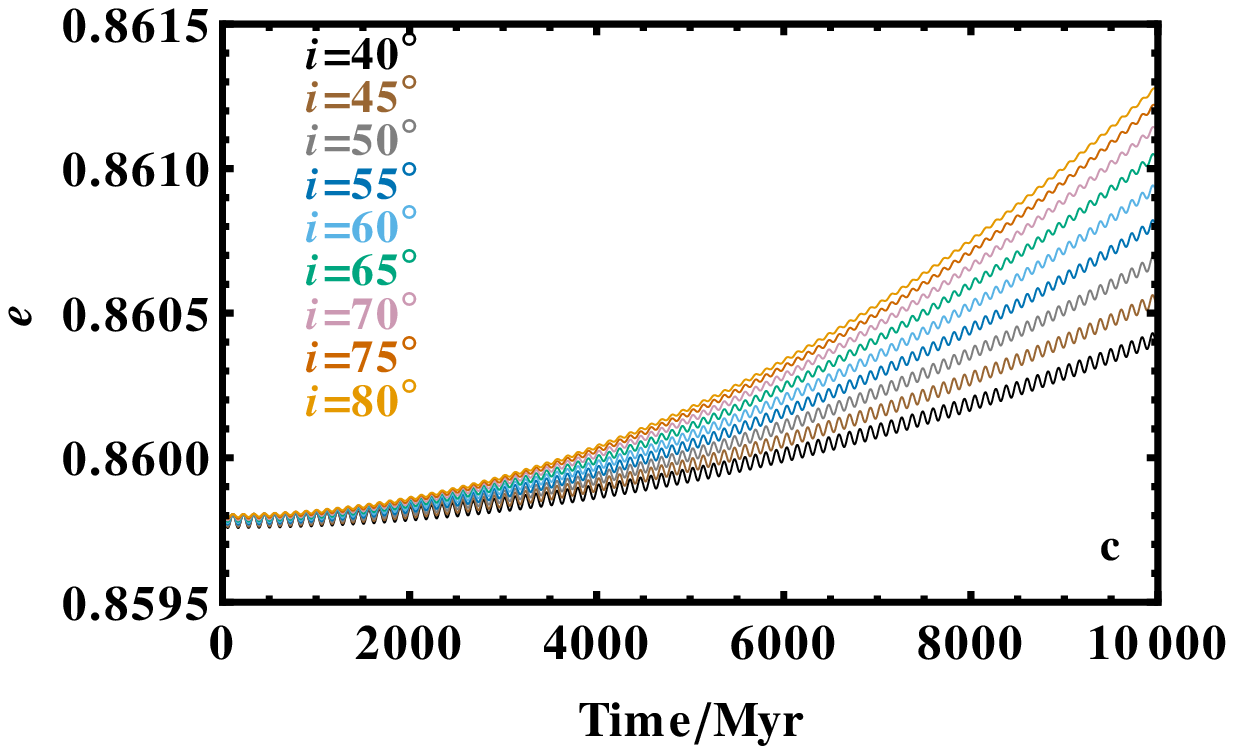}
}
\caption{The eccentricity evolution of the Sedna-like objects, all
with $a_0 = 544$ AU and $e_0 = 0.859$.  Panels a, b and c
correspond to ($\omega_0 = \omega_c$, $\Omega_0 = \Omega_c$),  
($\omega_0 = 225^{\circ}$, $\Omega_0 = 0^{\circ}$) and ($\omega_0 = \Omega_0 = 0^{\circ }$), respectively.
The modulation of the curves in panel c is due to planar tides.}
\label{FigSedna}   
\end{figure*}

\begin{figure*}
\centerline{\put(50,0){\LARGE {{\bf \ \ \ \ \ \ \ \ \ \ \ Scattered Disk Bodies}}}}
\centerline{
\includegraphics[width=0.70\textwidth,height=6cm]{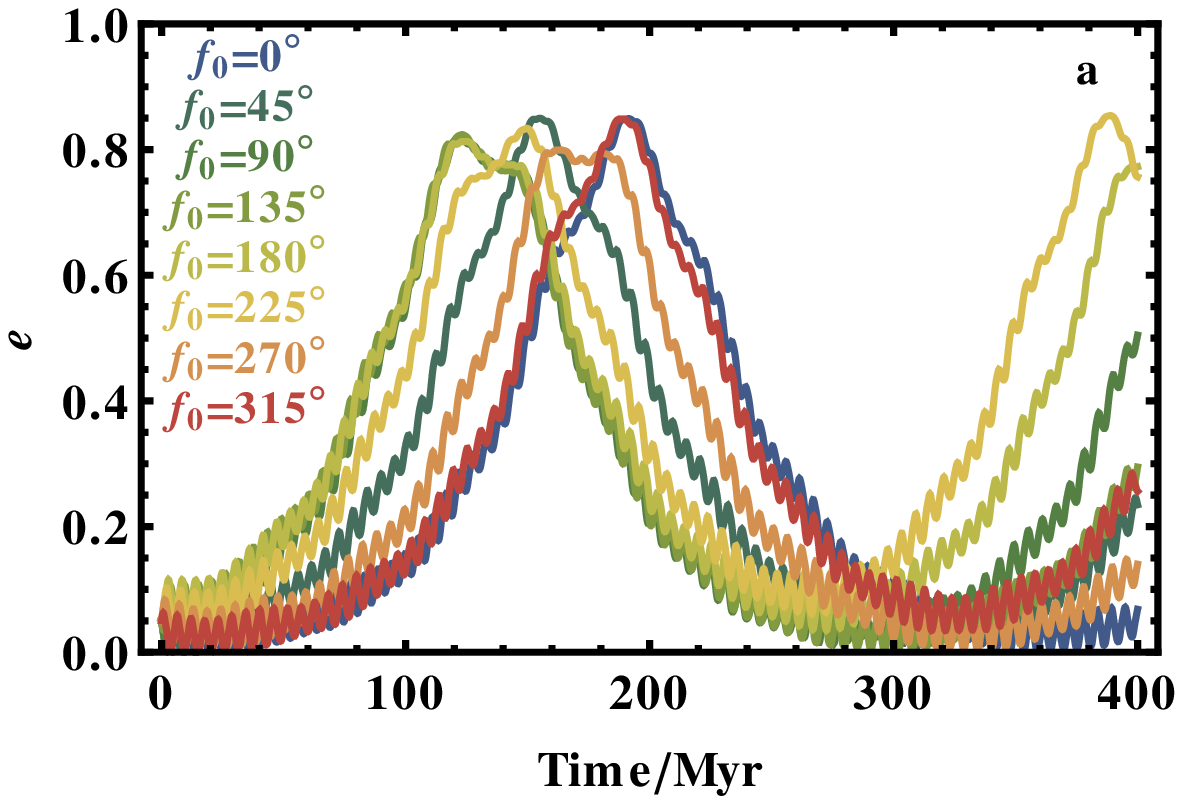}
}
\centerline{
\includegraphics[width=0.70\textwidth,height=6cm]{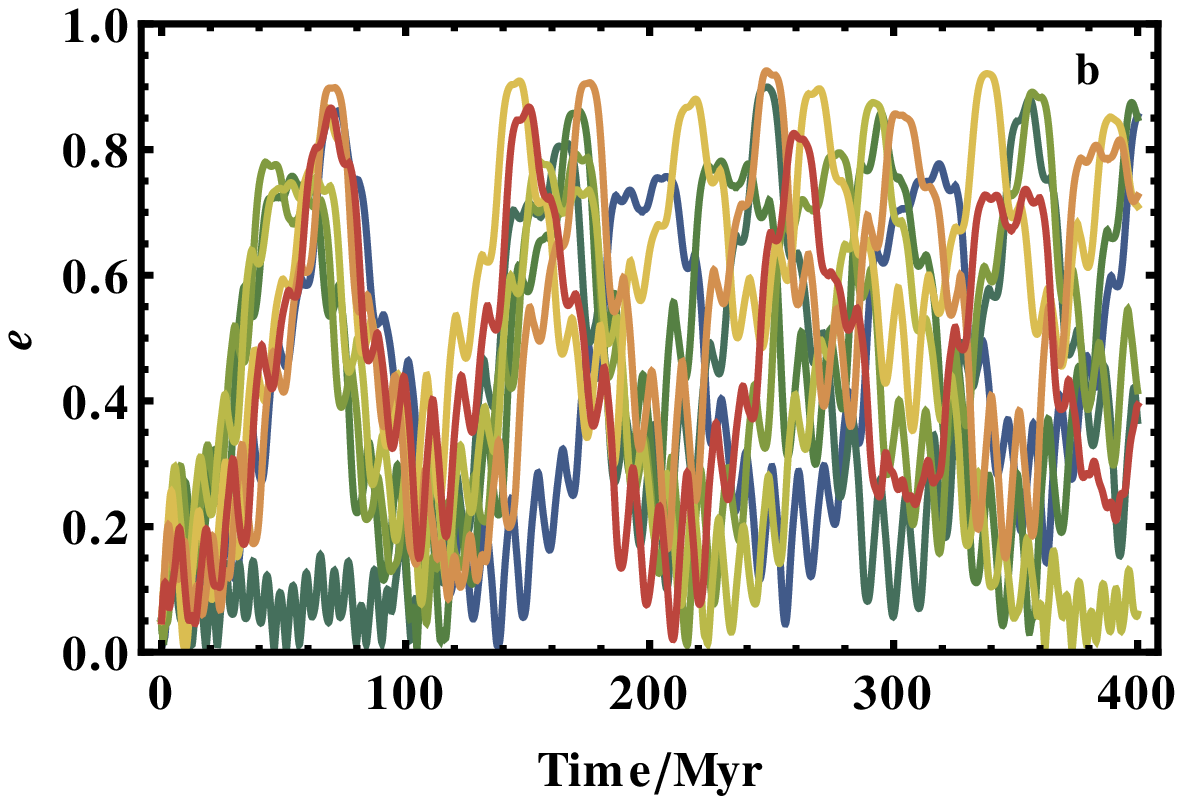}
}
\centerline{
\includegraphics[width=0.70\textwidth,height=6cm]{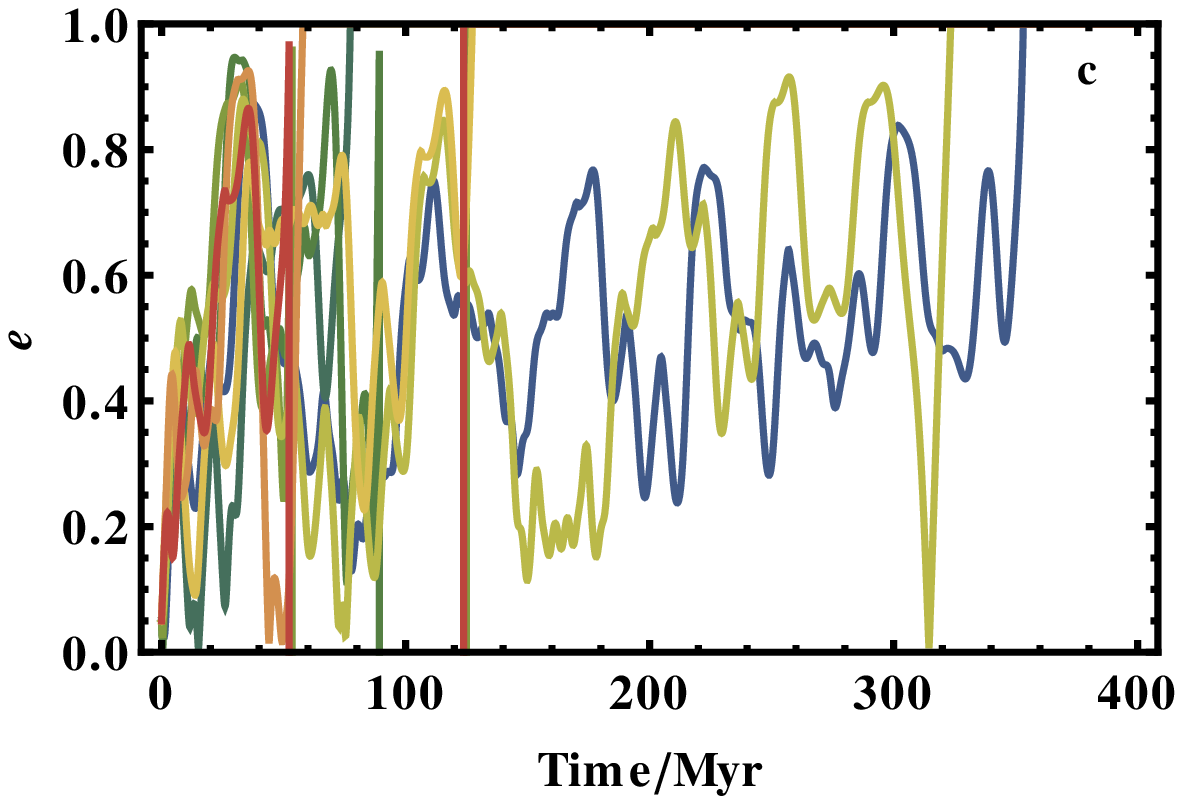}
}
\caption{The eccentricity evolution of initially nearly circular ($e_0 = 0.05$)
  scattered disk bodies (panel a: $a_0 = 3 \times
  10^4$ AU; panel b: $a_0 = 5 \times 10^4$ AU; panel c: $a_0 = 7 \times 10^4$ AU) 
  in systems inclined at $i = 60^{\circ}$ with respect to the
  Galactic plane.  The initial true anomalies of the objects are given
  in the legends in the upper panel, and $\omega_0 = \Omega_0 =
  0^{\circ}$ in all cases.  In the adiabatic regime, all curves on a
  given plot would be equivalent.  The plot demonstrates how
  adiabaticity is broken.}
\label{Fig2}   
\end{figure*}

\begin{figure*}
\centerline{\put(50,0){\LARGE {{\bf Planar Transition to
        Non-Adiabaticity}}}}
\centerline{ }
\
\centerline{ }
\centerline{\put(50,0){\LARGE {{\bf \ \  Prograde \ \ \ \ 
\ \ \ \ \ \ \ \ \ \ \ \ \ \ \ \ \ Retrograde}}}}
\centerline{
\includegraphics[width=0.50\textwidth,height=6cm]{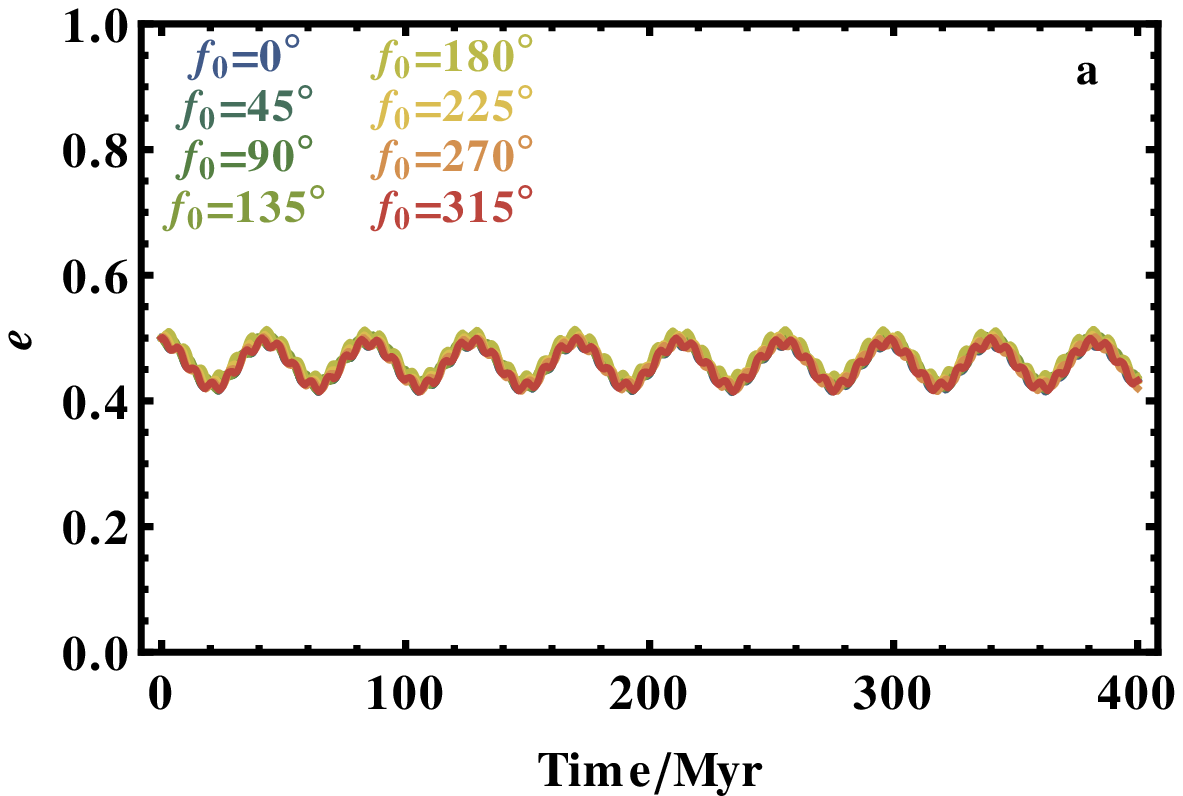}
\includegraphics[width=0.50\textwidth,height=6cm]{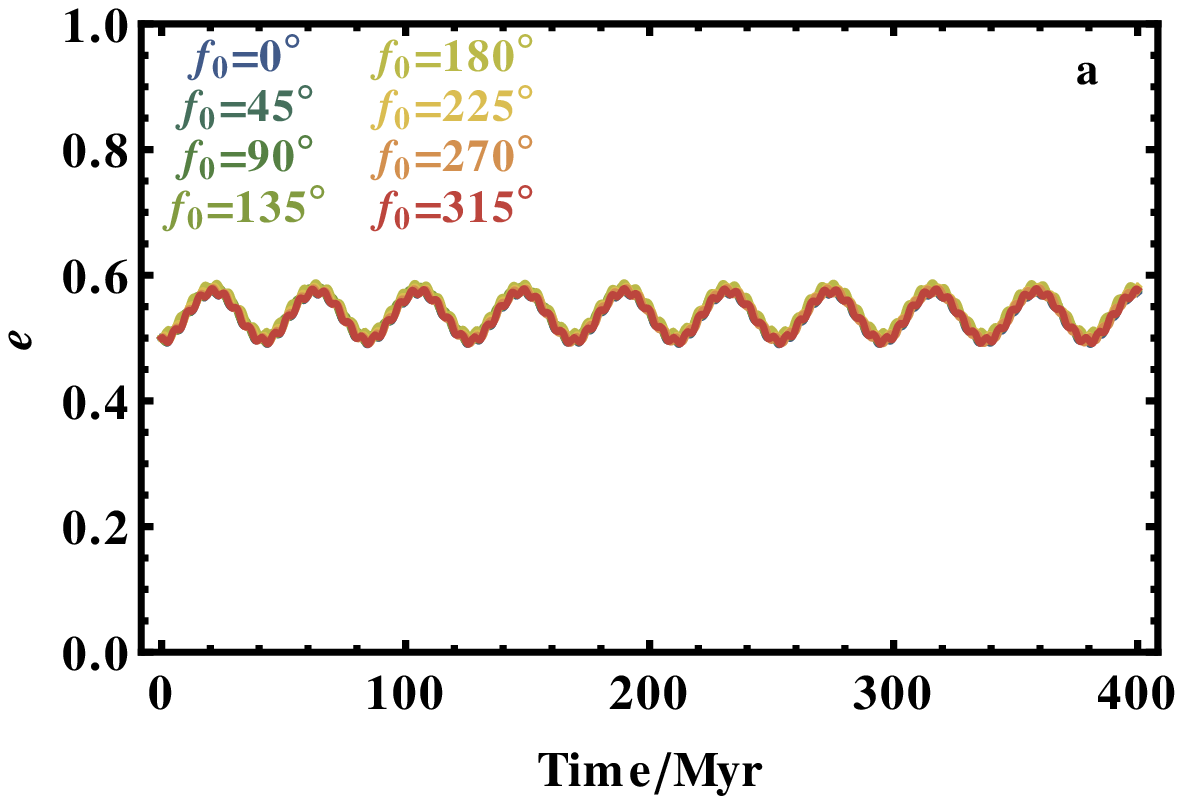}
}
\centerline{
\includegraphics[width=0.50\textwidth,height=6cm]{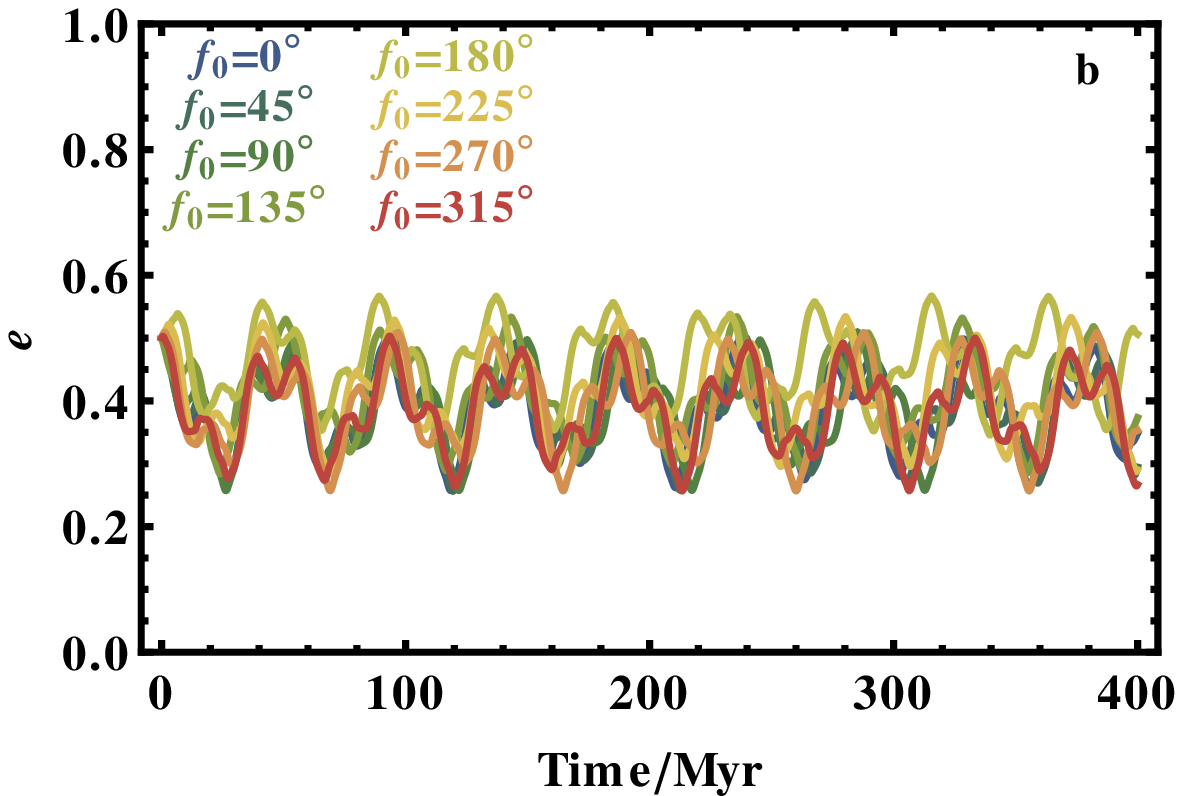}
\includegraphics[width=0.50\textwidth,height=6cm]{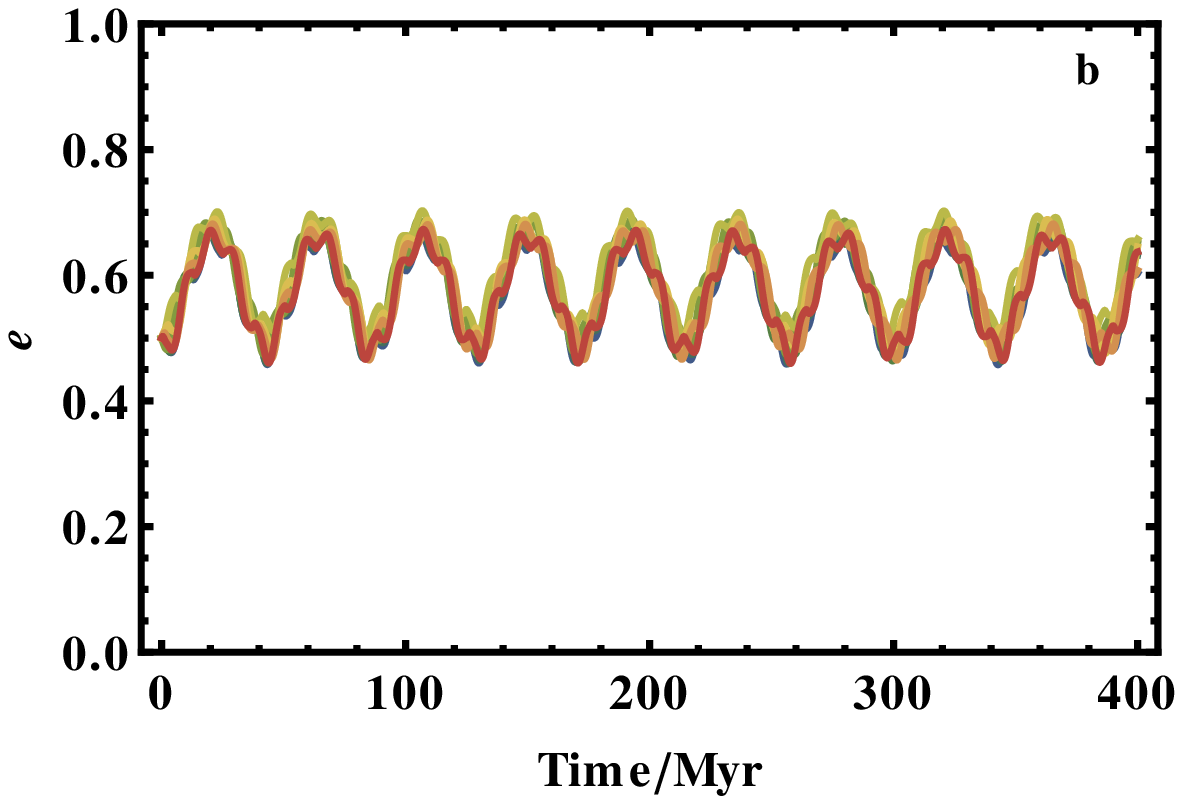}
}
\centerline{
\includegraphics[width=0.50\textwidth,height=6cm]{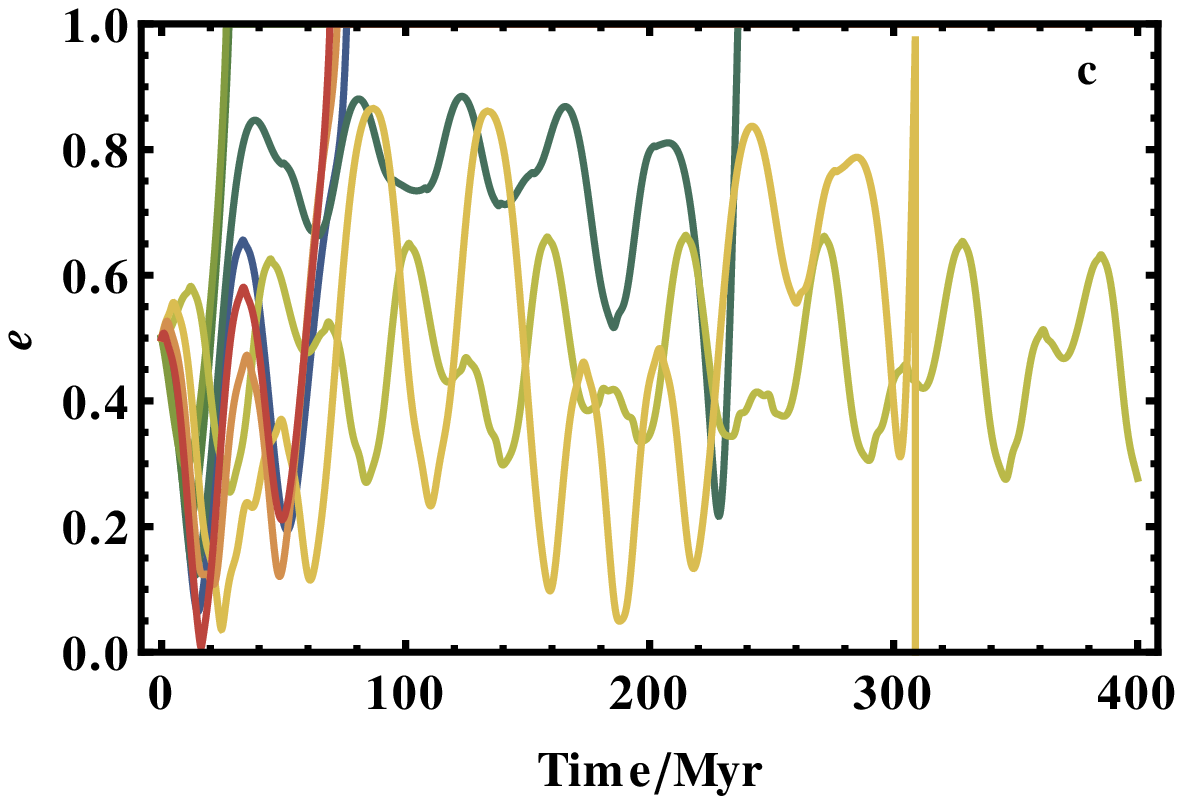}
\includegraphics[width=0.50\textwidth,height=6cm]{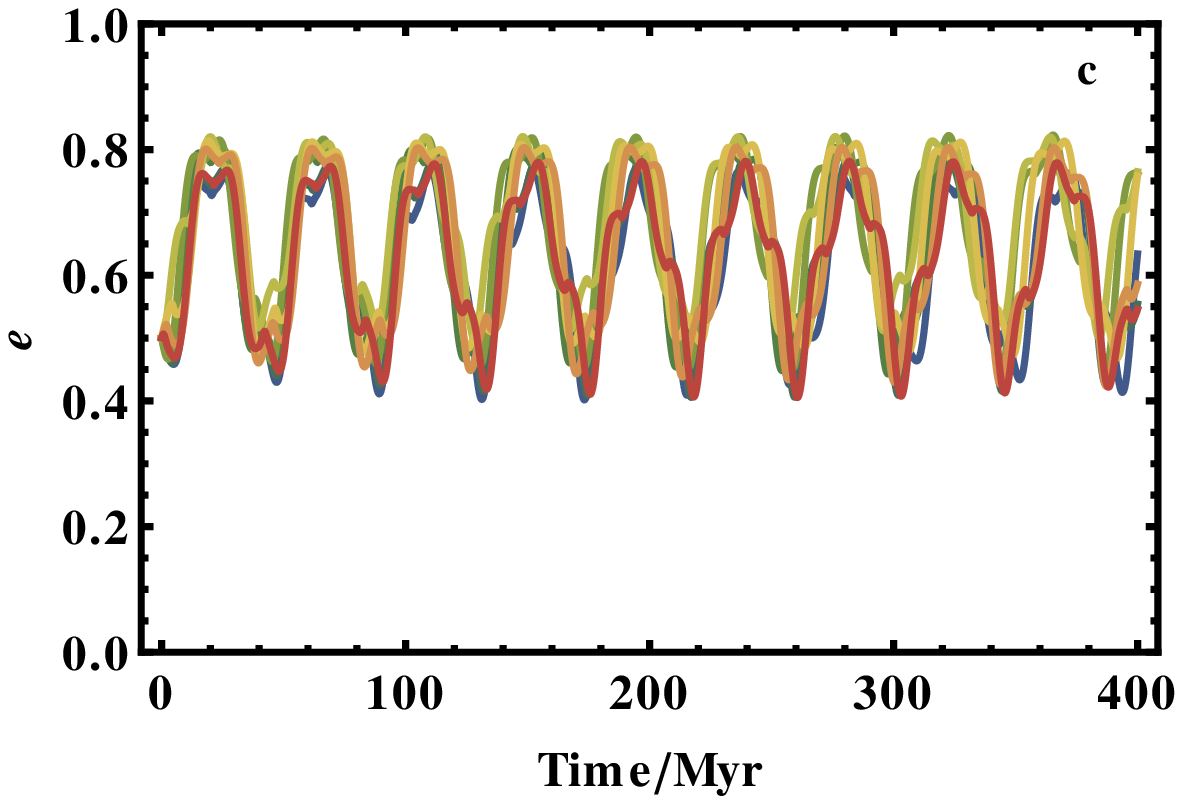}
}
\caption{A comparison of prograde planar and retrograde planar 
  transitions to non-adiabaticity.  Here, $i=0^{\circ}$ so vertical 
  tides vanish.  The left panels represent the prograde case 
  with $\varpi_0 = 0^{\circ}$ and the right panels represent the 
  retrograde case with $\varsigma = 0^{\circ}$. The semimajor axes 
  sampled here are equal to those in Fig. \ref{Fig2} 
  ($3 \times 10^4$ AU for the a panels, 
   $5 \times 10^4$ AU for the b panels, and
   $7 \times 10^4$ AU for the c panels).  The plot demonstrates
  how retrograde planets are generally more stable and more adiabatic 
  than their prograde counterparts. }
\label{Fig6}   
\end{figure*}


\subsection{Wide-orbit planet WD 0806-661B b}

\cite{luhetal2011} discovered the substellar companion to WD 0806-661B using
direct imaging.  This procedure yields a projected separation but no direct information about
the orbital properties, including the eccentricity.  Hence we assume the semimajor axis
is equal to the projected separation ($a_0 = a \approx 2500$ AU), and provide representative
values for other orbital properties: $e_0 = 0.5$, $\varpi_0 = \Omega_0 = 0^{\circ}$.  The
dependence of the orbital evolution on the mass of the planet is negligible.

In Fig. \ref{FigWideOrbit}, we plot the eccentricity evolution of WD 0806-661B b
for 9 different values of $i_0$ over 10 Gyr, a characteristic main sequence lifetime.
When $i_0 \gtrsim 42^{\circ}$, the eccentricity changes by over $0.1$.  When
$i_0 \gtrsim 71^{\circ}$, the eccentricity changes by over $0.2$.  These variations
illustrate that orbital signatures can significantly depend on the main sequence age 
of a wide-orbit planet.

\subsection{Sedna-like bodies}

Sedna represents the most distant member of the Solar System ever observed,
and currently resides about 90 AU from the Sun \citep{broetal2004}, close to
perihelion.  A recent orbital fit in terms of orbital elements for Sedna has been
computed by the Jet Propulsion 
Laboratory\footnote{As of September 26, 2012, from http://ssd.jpl.nasa.gov/sbdb.cgi?sstr=Sedna}
and shows that $a_c \approx 544$ AU, $e_c \approx 0.859$, $\omega_c \approx 310.9^{\circ}$
and $\Omega_c \approx 144.42^{\circ}$ where the subscript ``c'' stands for ``current''.
Sedna is inclined with respect to the ecliptic by about $12^{\circ}$.
The ability for Sedna to retain unperturbed by other Solar System objects until the end 
of the Sun's main sequence lifetime is uncertain; during the Sun's post-main sequence 
evolution, over 6 Gyr from now, Sedna will then evolve significantly \citep{verwya2012}.

Because of this uncertainty, and because the purpose of this paper is to 
present the general perturbed equations of motion, we model the evolution 
of several Sedna-like objects.  We fix $a_0 = a_c$ and $e_o = e_c$ but 
vary $i_0$, $\omega_0$ and $\Omega_0$, and run the 
simulations for 10 Gyr to show the full extent of the orbital change over a typical main
sequence lifetime.  We present our results in Fig. \ref{FigSedna}.  Panels a, b and c
correspond to ($\omega_0 = \omega_c$, $\Omega_0 = \Omega_c$),  
($\omega_0 = 225^{\circ}$, $\Omega_0 = 0^{\circ}$) and ($\omega_0 = \Omega_0 = 0^{\circ }$), respectively.
These sets of values were chosen to show different evolutionary behaviours.
In panel a, the eccentricity decreases by several hundredths; in panel b, the eccentricity
increases by several hundredths; in panel c, the eccentricity changes only by a few ten-thousandths.
The curves in panel c also exhibit a visually discernible modulation due to the planar
tides.

\subsection{The breaking of adiabaticity}

Here we demonstrate how adiabaticity is broken 
by considering outer scattered disk/inner Oort cloud-like objects 
with $a = 3 \times 10^{4}$ AU - $7 \times 10^{4}$ AU.
We integrate the general equations (Eqs. \ref{gena}-\ref{genf}) in 
Fig. \ref{Fig2}, as well as the planar prograde and retrograde 
limits in Fig. \ref{Fig6}.  

Figure \ref{Fig2} illustrates in panel a that the transition region between
adiabaticity and non-adiabaticity occurs just beyond about $3 \times 10^4$ AU.
In panel b, at $a_0 = 5 \times 10^4$ AU, the sinusoidal form of the eccentricity
evolution is heavily modulated, particularly after $200$ Myr have elapsed.  
In panel c, at $a_0 = 7 \times 10^4$ AU, all of the objects have become
unstable before $400$ Myr have elapsed.  Also 
note that the initially small eccentricity of the
orbiting bodies do not protect them from large eccentricity variations.

Because planar tides are weaker than vertical tides, the adiabatic
approximation should remain robust to larger distances in the planar
case than in the non-planar case (Fig. \ref{Fig2}).  Fig. \ref{Fig6}
demonstrates that the difference may be over a factor of
2 in semimajor axis the prograde case and over a factor of 3 in the
retrograde case.  Also, Fig. \ref{Fig6} demonstrates
that retrograde objects are more stable than prograde objects.  This
result is intuitively sensible, given that in the planar restricted
circular three-body problem, the Hill sphere is largest for a tertiary
orbiting the secondary in the opposite sense of the secondary's orbit
about the primary \citep[][pp. 131-133]{valkar2006}.  Thus, objects on
retrograde orbits are not likely to be as dynamically excited as
objects on prograde orbits, or ones with nonzero inclinations.

\section{Conclusions}

This work provides an analytical characterization of the perturbed
two-body problem.  In principle, the perturbations $\Upsilon$ may
represent any force; the position and velocity-dependent perturbations
we chose here are particularly well-suited for modelling the effect of
Galactic phenomena on single-planet exosystems.

The main results are the compact form (Eq. \ref{geneq}), from which
perturbative equations of motion may be derived analytically and
quickly using an algebraic software program\footnote{A {\it
    Mathematica} notebook with the necessary precomputed matrices is
  available from the authors.}, the general non-adiabatic non-planar
equations of motion in terms of orbital elements for several types of
perturbations (Eqs. \ref{gena}-\ref{genf}), and the
explicitly-expressed specific cases of these equations
(Eqs. \ref{adia} - \ref{adivarpi}, Eqs. \ref{ppaa}-\ref{ppaom}, and
Eqs. \ref{praa}-\ref{praom}).  We have discussed applications to 
wide-separation planets as well as distant trans-Neptunian objects in our 
Solar system.

Further, there are numerous applications of these equations in planetary
dynamics.  For example, a number of studies have investigated the idea
of a Galactic habitable zone~\citep{Gonz2001,Line2004}. These are the
regions in a galaxy in which a host star can harbor terrestrial
planets that can support life. Others have argued that perhaps the
entire Galaxy is (or has been) suitable for life, and the idea of
zones is too simplistic~\citep{Pran2008}.  There are many processes
influencing such a complex topic as habitability, including chemical,
geophysical and biological factors. However, a pre-requisite for life
is the existence of long-lived, stable planetary orbits around a host
star. In particular, the planet must remain largely unaffected by
perturbations from Galactic tides, spiral arms, bars and rings, giant
molecular clouds and dark matter substructures. The host star may lie
on a nearly circular orbit within the Galactic disk -- much like the
Sun -- or it may lie on a more eccentric orbit belonging to the
Galactic disk, spheroid or bulge. A substantial task for dynamicists
is to map out the regions of our Galaxy in which host stars can
provide planetary orbits that are stable against a rich variety of
perturbations. The formalism that we have developed in this paper 
can play an important role in this ambitious quest.

\section*{Acknowledgments}

We thank the referees for useful input which has improved the quality
of this paper.

\bibliographystyle{spbasic}

\appendix

\section{Some Auxiliary Variables}

Here, we collect some cumbersome variables referred to in the main body
of the paper. The variables $C_1, \dots C_9$ in the general case of
Eqs.~(\ref{gena})-(\ref{genf}) are given by
\begin{eqnarray}
C_1 &\equiv& e \cos{\omega} + \cos{\left(f+\omega\right)}
\nonumber
\\
C_2 &\equiv& e \sin{\omega} + \sin{\left(f+\omega\right)}
\nonumber
\\
C_3 &\equiv& \cos{i} \sin{\Omega} \sin{\left(f+\omega\right)}
                   - \cos{\Omega} \cos{\left(f+\omega\right)}
\nonumber
\\ 
C_4 &\equiv& \cos{i} \cos{\Omega} \sin{\left(f+\omega\right)}
                   + \sin{\Omega} \cos{\left(f+\omega\right)}
\nonumber
\\
C_5 &\equiv& \left(3 + 4e \cos{f} + \cos{2f} \right) \sin{\omega}
        + 2 \left( e + \cos{f} \right) \cos{\omega} \sin{f}
\nonumber
\\
C_6 &\equiv& \left(3 + 4e \cos{f} + \cos{2f} \right) \cos{\omega}
        - 2 \left( e + \cos{f} \right) \sin{\omega} \sin{f}
\nonumber
\\
C_7 &\equiv& \left(3 + 2e \cos{f} - \cos{2f} \right) \cos{\omega}
        + \sin{\omega} \sin{2f}
\nonumber
\\
C_8 &\equiv& \left(3  - \cos{2f} \right) \sin{\omega}
        - 2 \left(e + \cos{f} \right) \cos{\omega} \sin{f} 
\nonumber
\\
C_9 &\equiv& \left(3 + 2e \cos{f} - \cos{2f} \right) \sin{\omega}
        - \cos{\omega} \sin{2f}
\nonumber
\end{eqnarray}

The variables $C_{10}, \dots C_{15}$ in the adiabatic limit of
Eqs. (\ref{adia})-(\ref{adivarpi}) are given by
\begin{eqnarray}
C_{10} &\equiv& 5\left(e^2-2 \right) \sin{(2\omega)} \sin{(2\Omega)} \cos{i}
\nonumber
\\
C_{11} &\equiv& \cos{2\Omega} \left(1 - 6e^2 - 5 
\left(-3 + 2 e^2 \right) \cos{2\omega} - 10 \cos{2i} \sin^2{\omega}  \right)
\nonumber
\\
C_{12} &\equiv& 11 - 6e^2 + \cos{2\omega} \left(5 - 10e^2 \right)
+
10 \cos{2i}\sin^2{\omega}
\nonumber
\\ 
C_{13} &\equiv& \sin{2\Omega} 
\left(-1 + 6e^2 + 5\left(-3 + 2e^2\right) \cos{2\omega}
+
10 \cos{2i} \sin^2{\omega} \right)
\nonumber
\\
C_{14} &\equiv& \cos{2\Omega} \cos{i} \bigg[e^4 + \cos{2\omega}
\left(
e^4 + 4 + 2\sqrt{1-e^2} -2e^2 \left(3 + \sqrt{1-e^2}\right)
\right)
\bigg] 
\nonumber
\\
C_{15} &\equiv& \cos{2\Omega} \sin{2\omega} \left(3 + \cos{2i} \right)
\left( e^4 + 4 + 2 \sqrt{1-e^2}  - 6 e^2 - 2 e^2 \sqrt{1-e^2} \right) \nonumber
\end{eqnarray}

\end{document}